\documentclass[aps,pra,reprint,superscriptaddress,nofootinbib]{revtex4-2}

\usepackage{verbatim,amsmath,amssymb,color,appendix,algorithm,tensor,mathtools,mathrsfs,enumitem,graphicx}
\usepackage[caption=false,subrefformat=parens,labelformat=parens]{subfig}
\usepackage{hyperref}
\usepackage[all]{hypcap}    
\hypersetup{
  colorlinks=true,
  linkcolor=blue,
  citecolor=blue,
  urlcolor=blue,
  linkbordercolor={0 0 1}
}

\newcommand{\bra}[1]{\langle{#1}|}
\newcommand{\ket}[1]{|{#1}\rangle}
\newcommand{\braket}[1]{\langle{#1}\rangle}

\begin{document}


\title{Matrix Product States and Numerical Mode Decomposition for the Analysis of Gauge-Invariant Cavity Quantum Electrodynamics}


\author{Christopher J. Ryu}
\affiliation{Department of Electrical and Computer Engineering, University of Illinois Urbana-Champaign, Urbana, IL 61801, USA}


\author{Dong-Yeop Na}
\affiliation{Department of Electrical Engineering, Pohang University of Science and Technology, Pohang 37673, Republic of Korea}

\author{Weng C. Chew}
\email{wcchew@purdue.edu}
\affiliation{Department of Electrical and Computer Engineering, University of Illinois Urbana-Champaign, Urbana, IL 61801, USA}
\affiliation{Elmore Family School of Electrical and Computer Engineering, Purdue University, West Lafayette, IN 47907, USA}

\date{\today}

\begin{abstract}
There has been a problem of gauge ambiguities with the Rabi Hamiltonian due to the fact that it can be derived from two formally different but physically equivalent fundamental Hamiltonians. This problem has recently been resolved for models with single quantized electromagnetic mode. In this work, we mathematically and numerically verify this for multimode models. With this established, we combine the numerical methods, matrix product states (MPS) and numerical mode decomposition (NMD), for analyzing cavity QED systems. The MPS method is used to efficiently represent and time evolve a quantum state. However, since the coupling structure of the Rabi Hamiltonian is incompatible with MPS, it is numerically transformed into an equivalent Hamiltonian that has a chain coupling structure, which allows efficient application of MPS. The technique of NMD is used to extract the numerical electromagnetic modes of an arbitrary environment. As a proof of concept, this combined approach is demonstrated by analyzing 1D cavity QED systems in various settings.
\end{abstract}


\maketitle

\section{Introduction}
Quantum computers are becoming increasingly scalable and reliable \cite{Google_2019_Quantum_supremacy,Jian-Wei_Pan_2020_Quantum_computational_advantage,Jian-Wei_Pan_2021_Gaussian_boson_sampling,Xanadu_2022_Quantum_computational_advantage}, and cavity quantum electrodynamics (QED) in the ultrastrong coupling (USC) regime will potentially allow the operations to be incredibly fast \cite{Stassi_Cirio_&_Nori_2020_Scalable_QC_in_USC_regime} as a result of the field-atom coupling coefficient that is engineered to be comparable to the field frequency. The USC and even deep strong coupling regimes have been experimentally realized on superconducting circuits \cite{Niemczyk_et_al_2010_Circuit_QED_in_the_USC_regime,Yoshihara_et_al_2017_Superconducting_circuit_USC_regime,Wang_et_al_2020_Bloch-Siegert_shift_in_US_coupled_circuit_QED} and various other platforms \cite{Frisk_Kockum_2019_Ultrastrong_coupling}. This necessitates accurate numerical analysis of cavity QED systems in the USC regime, which is particularly challenging due to the sheer size of the Hilbert space when the atom is coupled to multiple electromagnetic modes of the cavity. In this work, we formulate a numerical analysis method for cavity QED, where a two-level atom (TLA) is placed in a complex electromagnetic environment.

The quantum Rabi model, which is based on the Rabi model \cite{Rabi_1936_Space_quantization} but with quantized electromagnetic fields, is suitable for studying a TLA interacting with the electromagnetic fields of the surrounding structure, e.g., a cavity. The model is beyond the rotating-wave approximation, which makes it valid even in the USC and deep strong coupling regimes of field-atom interactions \cite{Frisk_Kockum_2019_Ultrastrong_coupling}.

However, there has been a problem of gauge ambiguities due to two reasons: (i) there are two fundamental Hamiltonians different in their forms from which the Rabi Hamiltonians can be derived \cite{Leonardi1986}; and (ii) two-level truncation of the atomic Hilbert space may ruin gauge invariance \cite{Stokes_and_Nazir_2020_Gauge_non-invariance} if not applied properly. The two fundamental Hamiltonians are called the minimal coupling and electric dipole Hamiltonians \cite{Milonni}; or equivalently, $p\cdot A$ and $r\cdot E$ interactions \cite{Rzazewski2004,Loudon_2000_Quantum_theory_of_light} or Hamiltonians in the Coulomb gauge and dipole gauge \cite{Di_Stefano_et_al_2019_Resolution_of_gauge_ambiguities,Taylor_et_al_2020_Resolution_of_gauge_ambiguities}. Recently, the ambiguities have been resolved for models with single electromagnetic mode \cite{Di_Stefano_et_al_2019_Resolution_of_gauge_ambiguities,Taylor_et_al_2020_Resolution_of_gauge_ambiguities}. In these works, a proper way of applying two-level truncation to the atomic Hilbert space has been established, and from this, gauge invariant Rabi Hamiltonians have been derived.

It has been pointed out by Mu\~noz \textit{et al}.\ \cite{Munoz_superluminal_2018} that the quantum Rabi model in the USC regime and beyond may violate relativistic causality, and to avoid this, multiple field modes of the cavity (in which the TLA is placed) must be taken into account. In \cite{Munoz_superluminal_2018}, this has been numerically demonstrated using the matrix product state (MPS) method, where periodic boundary conditions (PBC) have been employed with the TLA placed at the center of a 1D lattice. Similar work has been done by Flick \textit{et al}.\ \cite{Flick2017} in the context of 1D cavity QED, where they consider a truncation of the higher energy part of the Hilbert space. In particular, bosonic Fock states with a total number of photons beyond two were truncated.

Our work is focused on formulating a generalized simulation strategy and mainly consists of the following three contributions.
\begin{enumerate}[label=\arabic*),topsep=0pt,itemsep=-1ex,partopsep=1ex,parsep=1ex]
\item We identify the gauge invariant multimode Rabi Hamiltonians, where both the two-level truncation of the atom and proper truncation of the electromagnetic modes \cite{Taylor_et_al_2022_Mode_truncation_cQED} are considered. We confirm that previous conclusions regarding gauge invariance of single mode Rabi Hamiltonians in \cite{Di_Stefano_et_al_2019_Resolution_of_gauge_ambiguities,Taylor_et_al_2020_Resolution_of_gauge_ambiguities} extend to the multimode Rabi Hamiltonians. In particular, we compute and compare the energy eigenvalue spectra of various multimode Hamiltonians.
\item We stabilize the numerical transformation scheme that turns the multimode Rabi Hamiltonian into the chain Hamiltonian. This allows transformations of not just linearly distributed electromagnetic mode frequencies but also arbitrarily distributed frequencies.
\item We present a formulation that combines MPS with numerical mode decomposition (NMD) \cite{Qinfo_preserving_CEM,Na_Zhu_Chew_2021}, which is a computational electromagnetics (CEM) technique of solving classical Maxwell's equations, to simulate arbitrary, inhomogeneous cavity QED settings.
\end{enumerate}

\section{Gauge Invariance of Multimode Rabi Hamiltonians}
The problem of gauge ambiguities of single mode quantum Rabi models has recently been resolved \cite{Di_Stefano_et_al_2019_Resolution_of_gauge_ambiguities,Taylor_et_al_2020_Resolution_of_gauge_ambiguities}. The proper truncation of electromagnetic modes without considering material truncation (two-level truncation) has been studied \cite{Taylor_et_al_2022_Mode_truncation_cQED}. In this work, we consider both the two-level truncation of the atom and the proper truncation of electromagnetic modes. We present the relevant cavity QED Hamiltonians in this section and verify their gauge invariance properties. For detailed derivations of the Hamiltonians presented in this section, see Appendices~\ref{app:min_coup_H} through \ref{app:trunc_method_and_gauge_invar}.

\subsection{Relevant Hamiltonians}\label{sec:relevant_Hamiltonians}
\subsubsection*{Fundamental Hamiltonians}
The multimode Rabi Hamiltonians can be derived from two formally different but physically equivalent fundamental Hamiltonians. One of them is the Hamiltonian in the Coulomb gauge,
\begin{equation}
\hat{H}_C=\frac{[\hat{\bf p}-q\hat{\bf A}({\bf r}_0)]^2}{2m}+V(\hat{\bf r})+\hat{H}_F,\label{eq:H_C}
\end{equation}
where $\hat{\bf r}$ and $\hat{\bf p}$ are the canonically conjugate position and momentum operators for the moving charge in the atom; $m$ and $q$ are its mass and charge; $\hat{\bf A}$ is the vector potential operator; ${\bf r}_0$ is the position of the nucleus; $V(\hat{\bf r})=q\Phi(\hat{\bf r})$ is the scalar potential that binds the free charge; and the free field Hamiltonian is
\begin{equation}
\hat{H}_F=\int dV\left(\frac{\hat{\boldsymbol{\Pi}}^2({\bf r})}{\epsilon({\bf r})}+\frac{[\nabla\times\hat{\bf A}({\bf r})]^2}{\mu_0}\right)=\sum_{k=1}^{M}\hbar\omega_k\hat{a}_k^\dagger\hat{a}_k.
\end{equation}
In the above, $\epsilon({\bf r})$ is the permittivity that describes the medium inhomogeneity inside the cavity, $\mu_0$ is the permeability of free space, $\hat{\boldsymbol{\Pi}}({\bf r})=\epsilon({\bf r})\partial\hat{\bf A}({\bf r})/\partial t$ is the canonically conjugate variable of $\hat{\bf A}$ \cite{QEM_a_new_look_I,QEM_a_new_look_II}, $M$ is the truncated number of cavity modes, $\omega_k$ is the mode frequency, and $\hat{a}_k^\dagger$ ($\hat{a}_k$) is the creation (annihilation) operator for mode $k$. The Hamiltonian (\ref{eq:H_C}) is in the Coulomb gauge ($\nabla\cdot\hat{\bf A}=0$) where the vector potential operator is purely transverse.

The Coulomb gauge Hamiltonian (\ref{eq:H_C}) can be transformed into a electric dipole interaction based Hamiltonian by the Power-Zienau-Woolley transformation \cite{Power_and_Zienau_1959_Coulomb_gauge_QED,Woolley_1971_Molecular_QED}, which is implemented with the unitary operator, $\hat{U}_{PZW}=e^{-iq\hat{\bf r}\cdot\hat{\bf A}({\bf r}_0)/\hbar}$. The dipole gauge Hamiltonian is obtained as
\begin{equation}\label{eq:PZW_transform}
\hat{H}_D=\hat{U}_{PZW}\hat{H}_C\hat{U}_{PZW}^\dagger,
\end{equation}
which is why the two fundamental Hamiltonians are physically equivalent \cite{Cohen-Tannoudji_Photons_and_Atoms}, and it is written out:
\begin{equation}
\hat{H}_D=\frac{\hat{\bf p}^2}{2m}+V(\hat{\bf r})-\hat{\bf d}\cdot\hat{\bf E}_\perp({\bf r}_0)+\hat{H}_F+\sum_{k=1}^{M}\tfrac{[\hat{\bf d}\cdot{\bf A}_k({\bf r}_0)]^2}{2\epsilon_0 V_0},\label{eq:H_D}
\end{equation}
where $\hat{\bf d}=q\hat{\bf r}$ is the dipole moment operator; $\hat{\bf E}_\perp=-\partial\hat{\bf A}/\partial t$ is the transverse electric field operator; ${\bf A}_k$ is the vector potential spatial eigenfunction obtained by solving classical Maxwell's equations; $\epsilon_0$ is the permittivity of free space; and $V_0$ is the volume of the cavity that the atom is in.

\subsubsection*{Traditional Rabi Hamiltonians}
The Rabi Hamiltonians are obtained by applying two-level truncation to the atomic Hilbert space as an approximation. The projection operator, $\hat{\mathcal{P}}=\ket{g}\bra{g}+\ket{e}\bra{e}$, is used for this purpose, where $\ket{g}$ and $\ket{e}$ represent the ground and first excited states, respectively, of the bare atomic Hamiltonian $\hat{H}_A=\frac{\hat{\bf p}^2}{2m}+V(\hat{\bf r})$.

Traditionally, the Rabi Hamiltonians were obtained by a direct truncation of the Hamiltonian:
\begin{equation}\label{eq:TLT}
\hat{\mathcal{H}}_i'=\hat{\mathcal{P}}\hat{H}_i\hat{\mathcal{P}},
\end{equation}
where $i=C$ or $D$ for Coulomb or dipole gauge, and the prime indicates the direct truncation of the atomic Hilbert space. Calligraphic symbols with hats are used in this paper to represent operators in the two-level truncated atomic Hilbert space.

\begin{figure*}
\includegraphics[width=1\linewidth]{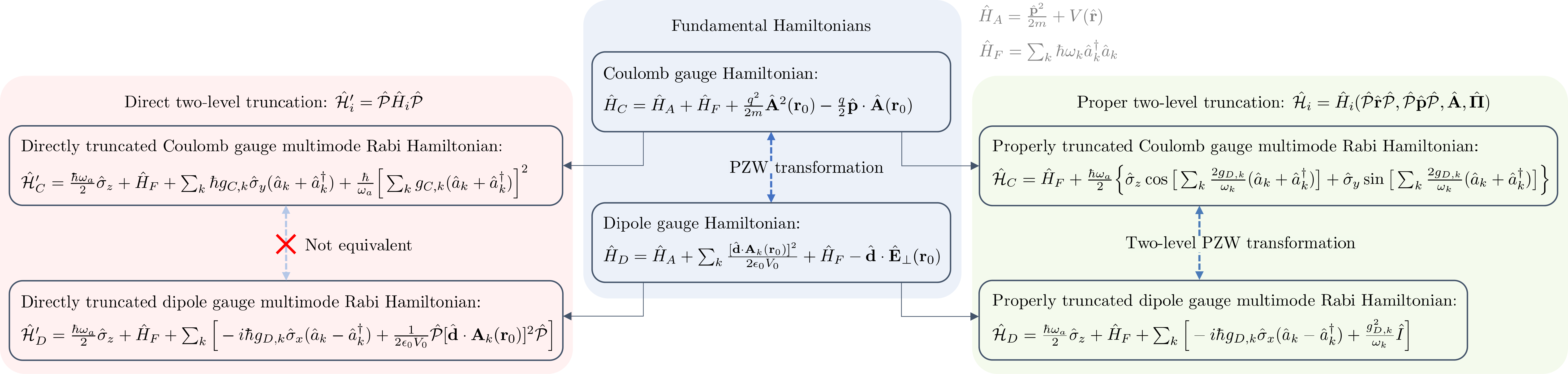}
\caption{The truncated Rabi Hamiltonians are derived from the fundamental Hamiltonians that are equivalent. The direct two-level truncation leads to Rabi Hamiltonians that are not equivalent anymore, but the proper two-level truncation preserves the unitarity relation.}
\label{fig:Ham_chart}
\end{figure*}

When the direct truncation is applied to (\ref{eq:H_C}), the resulting multimode Rabi Hamiltonian is
\begin{align}
\hat{\mathcal{H}}_C'&=\frac{\hbar\omega_a}{2}\hat\sigma_z+\sum_{k=1}^{M}\Big[\hbar\omega_k\hat{a}_k^\dagger\hat{a}_k+\hbar g_{C,k}\hat\sigma_y(\hat{a}_k+\hat{a}_k^\dagger)\Big]\notag\\
&\quad+\frac{\hbar}{\omega_a}\Big[\sum_{k=1}^{M}g_{C,k}(\hat{a}_k+\hat{a}_k^\dagger)\Big]^2,\label{eq:Rabi_H_C'}
\end{align}
where $\hbar\omega_a$ is the energy gap between $\ket{e}$ and $\ket{g}$ of the atom, and the mode-dependent coupling coefficient in the Coulomb gauge is
\begin{equation}\label{eq:g_C}
g_{C,k}=\frac{\omega_a{\bf d}\cdot{\bf A}_k({\bf r}_0)}{\sqrt{2\hbar\omega_k \epsilon_0 V_0}}
\end{equation}
where ${\bf d}=\braket{g|\hat{\bf d}|e}$ is the dipole moment vector assumed to be real \cite{DABMiller}. The coefficient of the diamagnetic term [last term in (\ref{eq:Rabi_H_C'})] is determined by the Thomas-Reiche-Khun (TRK) sum rule \cite{Reiche_and_Thomas_1925_Sum_rule,Kuhn_1925_Sum_rule,Wang_1999_Generalization_of_TRK_sum_rule,Tufarelli_et_al2015_A2_term}. Similarly, the multimode dipole gauge Rabi Hamiltonian is derived to be
\begin{align}
\hat{\mathcal{H}}_D'=\tfrac{\hbar\omega_a}{2}\hat\sigma_z+\sum_{k=1}^{M}\Big[\hbar\omega_k\hat{a}_k^\dagger\hat{a}_k-i\hbar g_{D,k}\hat\sigma_x(\hat{a}_k-\hat{a}_k^\dagger)&\notag\\
+\tfrac{1}{2\epsilon_0 V_0}\hat{\mathcal{P}}[\hat{\bf d}\cdot{\bf A}_k({\bf r}_0)]^2\hat{\mathcal{P}}&\Big]\label{eq:Rabi_H_D'}
\end{align}
with the coupling coefficient in the dipole gauge,
\begin{equation}\label{eq:g_D}
g_{D,k}={\bf d}\cdot{\bf A}_k({\bf r}_0)\sqrt{\tfrac{\omega_k}{2\hbar\epsilon_0 V_0}}.
\end{equation}
The coupling coefficients in the two gauges, (\ref{eq:g_C}) and (\ref{eq:g_D}), are related as $g_{C,k}=g_{D,k}\omega_a/\omega_k$. This is why for single electromagnetic mode that is resonant with the TLA, the interaction Hamiltonians of $\hat{\mathcal{H}}_C'$ and $\hat{\mathcal{H}}_D'$ appear equivalent \cite{Scully_and_Zubairy_1997_QO}.

However, there is no unitary operator that links the two directly truncated Rabi Hamiltonians, (\ref{eq:Rabi_H_C'}) and (\ref{eq:Rabi_H_D'}). Upon truncation, we obtain $\hat{\mathcal{H}}_C'=\hat{\mathcal{P}}\hat{H}_C\hat{\mathcal{P}}$ and $\hat{\mathcal{H}}_D'=\hat{\mathcal{P}}\hat{H}_D\hat{\mathcal{P}}=\hat{\mathcal{P}}\hat{U}_{PZW}\hat{H}_C\hat{U}_{PZW}^\dagger\hat{\mathcal{P}}$ using (\ref{eq:PZW_transform}) and (\ref{eq:TLT}). Although $\hat{H}_C$ and $\hat{H}_D$ are unitary transforms of each other as shown in (\ref{eq:PZW_transform}), $\hat{\mathcal{H}}_C'$ and $\hat{\mathcal{H}}_D'$ are not because $\hat{\mathcal{P}}\hat{U}_{PZW}$ is not unitary \cite{Di_Stefano_et_al_2019_Resolution_of_gauge_ambiguities,Taylor_et_al_2020_Resolution_of_gauge_ambiguities}.\footnote{For more mathematical details, see Appendix~\ref{app:trunc_method_and_gauge_invar}.} Therefore, gauge invariance has been lost in the process of direct two-level truncation.

Also, the last term in (\ref{eq:Rabi_H_D'}) is the dipole self-energy term that cannot be written in terms of the two-level Pauli operators. Due to the direct truncation method applied to the Hamiltonian, all eigenstates (and not just $\ket{g}$ and $\ket{e}$) of the atom are needed to express this term \cite{Taylor_et_al_2020_Resolution_of_gauge_ambiguities}.

\subsubsection*{Properly Truncated Rabi Hamiltonians}
The full Hamiltonian in either gauge [(\ref{eq:H_C}) or (\ref{eq:H_D})] can be thought of as a function of four conjugate operators: $\hat{H}_i(\hat{\bf r},\hat{\bf p},\hat{\bf A},\hat{\boldsymbol{\Pi}})$ with $i=C$ or $D$. The proper way to truncate the atomic part of these Hamiltonians is \cite{Di_Stefano_et_al_2019_Resolution_of_gauge_ambiguities,Taylor_et_al_2020_Resolution_of_gauge_ambiguities}
\begin{equation}\label{eq:proper_truncation}
\hat{\mathcal{H}}_i=\hat{H}_i(\hat{\mathcal{P}}\hat{\bf r}\hat{\mathcal{P}},\hat{\mathcal{P}}\hat{\bf p}\hat{\mathcal{P}},\hat{\bf A},\hat{\boldsymbol{\Pi}}).
\end{equation}
The difference between this and the direct truncation approach may seem subtle, but (\ref{eq:proper_truncation}) leads to a very different form of the Rabi Hamiltonian in the Coulomb gauge. In this gauge, the multimode Rabi Hamiltonian is obtained to be
\begin{align}\label{eq:Rabi_H_C}
\hat{\mathcal{H}}_C=\sum_{k=1}^{M}\hbar\omega_k\hat{a}_k^\dagger\hat{a}_k+\tfrac{\hbar\omega_a}{2}\Big\{\hat\sigma_z\cos\big[\textstyle\sum_{k=1}^{M}\tfrac{2g_{D,k}}{\omega_k}(\hat{a}_k+\hat{a}_k^\dagger)\big]\notag\\
+\hat\sigma_y\sin\big[\textstyle\sum_{k=1}^{M}\tfrac{2g_{D,k}}{\omega_k}(\hat{a}_k+\hat{a}_k^\dagger)\big]\Big\}.
\end{align}
Likewise, the one in the dipole gauge is
\begin{equation}\label{eq:Rabi_H_D}
\hat{\mathcal{H}}_D=\tfrac{\hbar\omega_a}{2}\hat\sigma_z+\hbar\sum_{k=1}^{M}\Big[\omega_k\hat{a}_k^\dagger\hat{a}_k-i g_{D,k}\hat\sigma_x(\hat{a}_k-\hat{a}_k^\dagger)+\tfrac{g_{D,k}^2}{\omega_k}\hat{\mathcal{I}}\Big]
\end{equation}
where $\hat{\mathcal{I}}=\hat{\mathcal{P}}$ is the identity operator in the two-level subspace of the atomic Hilbert space. The coefficient of the dipole self-energy term [last term in (\ref{eq:Rabi_H_D})] can be confirmed using the TRK sum rule for interacting photons \cite{Savasta_et_al_2021_TRK_sum_rule} (discussed in Appendix~\ref{app:TRK_sum_rule}). The properly truncated Rabi Hamiltonians, (\ref{eq:Rabi_H_C}) and (\ref{eq:Rabi_H_D}), are still unitarily related by the two-level PZW transformation: $\hat{\mathcal{H}}_D=\hat{\mathcal{U}}_{PZW}\hat{\mathcal{H}}_C\hat{\mathcal{U}}_{PZW}^\dagger$ where $\hat{\mathcal{U}}_{PZW}=e^{-iq\hat{\mathcal{P}}\hat{\bf r}\hat{\mathcal{P}}\cdot\hat{\bf A}({\bf r}_0)/\hbar}$. Thus, gauge invariance is preserved by the proper two-level truncation (\ref{eq:proper_truncation}). The various Hamiltonians presented in this section are summarized in Fig.\ \ref{fig:Ham_chart}.

\subsection{Spectra Comparison}
Although it is mathematically shown in Appendix~\ref{app:trunc_method_and_gauge_invar} that the properly truncated Rabi Hamiltonians [(\ref{eq:Rabi_H_C}) and (\ref{eq:Rabi_H_D})] are equivalent with each other, it would be interesting to see if they are good approximations of the fundamental Hamiltonians [(\ref{eq:H_C}) and (\ref{eq:H_D})]. To this end, the energy eigenvalue spectra of various Hamiltonians discussed in this section are numerically calculated and compared.

\begin{figure}
\includegraphics[width=.975\linewidth]{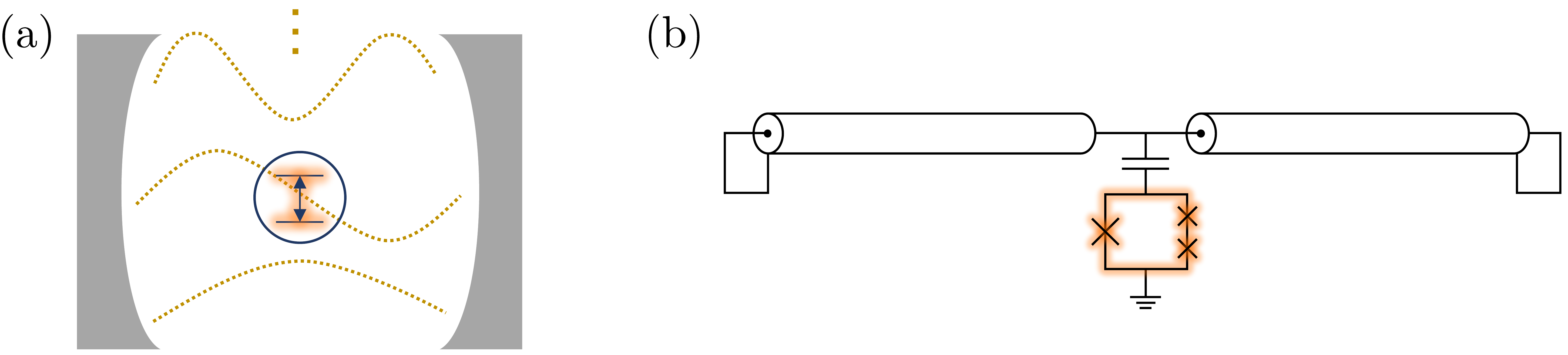}
\caption{(a) 1D simulation setting with a TLA placed at the center of a uniform PEC cavity. (b) Circuit analogue of (a) that involves a flux qubit capacitively coupled to a transmission line resonator.}
\label{fig:TLA_in_a_cavity}
\end{figure}

\begin{figure}
\includegraphics[width=1\linewidth]{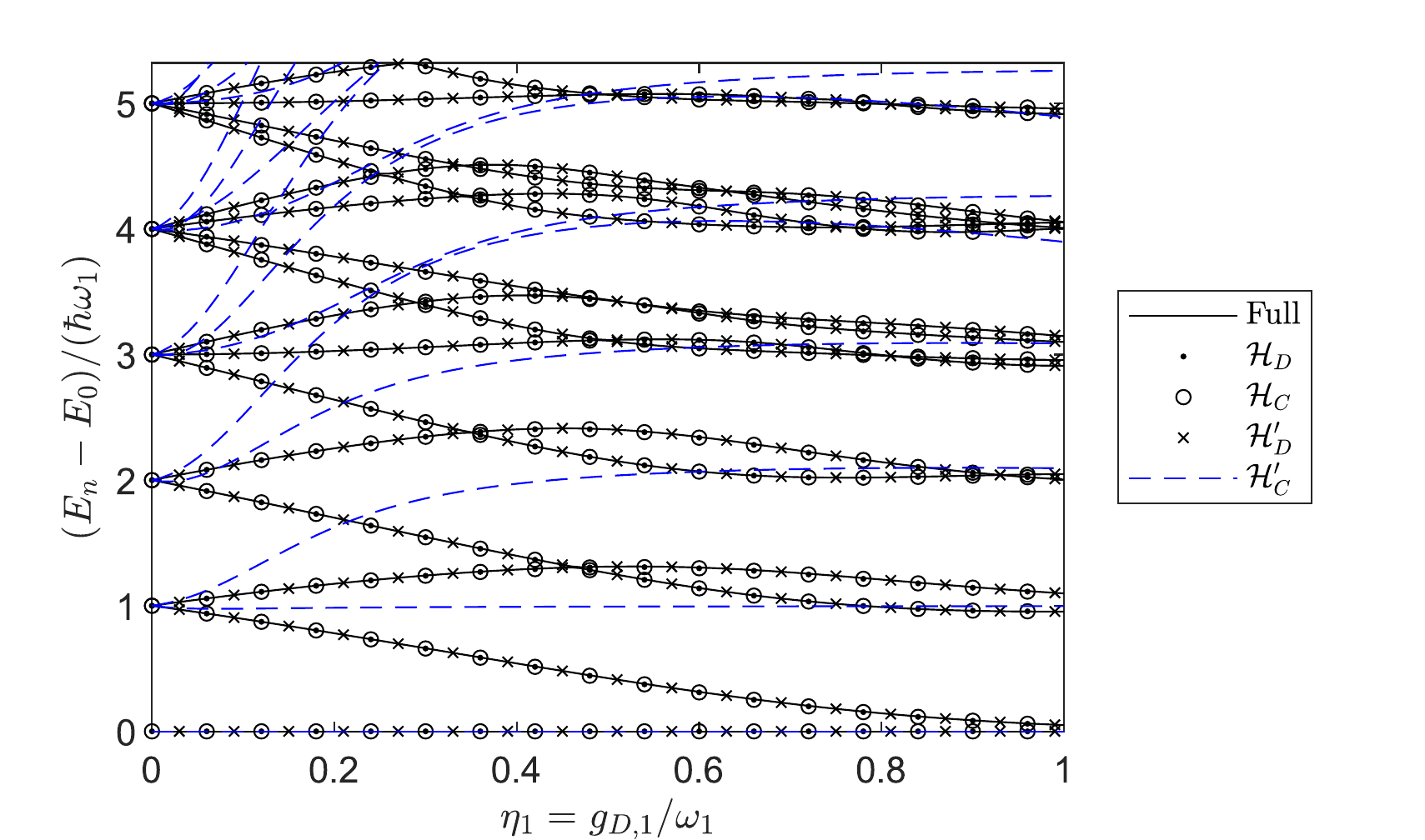}
\caption{Energy eigenvalue spectra of various multimode field-atom Hamiltonians derived in Sec.\ \ref{sec:relevant_Hamiltonians}. Vertical axis represents energy difference from the ground state normalized by the photonic energy of the fundamental mode of the cavity, and the horizontal axis is the normalized coupling strength of the same mode.}
\label{fig:3-mode_spectra}
\end{figure}

For simplicity, the atom is assumed to be placed at the center of a 1D perfect electric conductor (PEC) cavity, whose fundamental mode is resonant with the TLA at $\omega_a$. The setting is illustrated in Fig.\ \ref{fig:TLA_in_a_cavity}. The mode frequencies of the cavity are $\omega_k=k\omega_a$ with integer $k\in[1,M]$.\footnote{Assuming a 1D PEC cavity results in this simple and convenient situation. However, we emphasize that the conclusions drawn from the plot in Fig.~\ref{fig:3-mode_spectra} would not change even if we used a 3D cavity that has unevenly distributed electromagnetic mode frequencies.} A total of $M=5$ modes are considered for the calculation of the energy eigenvalue spectra shown in Fig.\ \ref{fig:3-mode_spectra}, but only the modes with odd index $k$ are coupled to the TLA due to its position. Using a larger $M$ would not make a difference in Fig.\ \ref{fig:3-mode_spectra} since the higher modes would only contribute to the higher part of the spectra that is not shown. For the full Hamiltonian, the atom is not truncated, and a double well potential is used to implement an anharmonic atom as done in \cite{De_Bernardis_et_al_2018_Breakdown_of_gauge_invariance,Stokes_and_Nazir_2020_Gauge_non-invariance}. To model the full atom numerically, the Fourier grid Hamiltonian approach is used \cite{Marston_and_Balint-Kurti_1989_Fourier_grid_Hamiltonian}.

The various multimode spectra in Fig.\ \ref{fig:3-mode_spectra} essentially show that except for $\hat{\mathcal{H}}_C'$ (\ref{eq:Rabi_H_C'}), the Hamiltonians agree with one another even when multiple electromagnetic modes are considered with two-level truncation. Because the double well potential used to model the full atom is highly anharmonic, very good agreements are observed among the various multimode Rabi Hamiltonians with the full Hamiltonian. Therefore, we confirm that the conclusions made about the single mode Rabi Hamiltonians \cite{Di_Stefano_et_al_2019_Resolution_of_gauge_ambiguities,Taylor_et_al_2020_Resolution_of_gauge_ambiguities} extend to the multimode case. From this point and onward, we primarily use the multimode Rabi Hamiltonian in the dipole gauge (\ref{eq:Rabi_H_D}) with the dipole self-energy term dropped. Note that because this term is proportional to an identity operator, dropping it does not make a difference to the eigenvalue spacings plotted in Fig.\ \ref{fig:3-mode_spectra}.

One last issue that should be touched upon when dealing with multimode cavity QED is the divergence of Lamb shifts when an infinite number of modes is considered without a cutoff. It has recently been shown that finite expressions can be obtained in circuit QED when gauge invariance is respected \cite{Malekakhlagh_et_al_2017_Cutoff-free_cQED} and that divergences can be avoided by rescaling the bare atomic parameters from circuit analysis \cite{Gely2017}. In this paper, we make a simplifying assumption that the TLA frequency is experimentally measured so that its energy levels do not depend on the number of modes considered \cite{Munoz_superluminal_2018}.

\section{Stable Numerical Transformation to the Chain Hamiltonian}\label{sec:chain_mapping}
\begin{figure}
\includegraphics[width=1\linewidth]{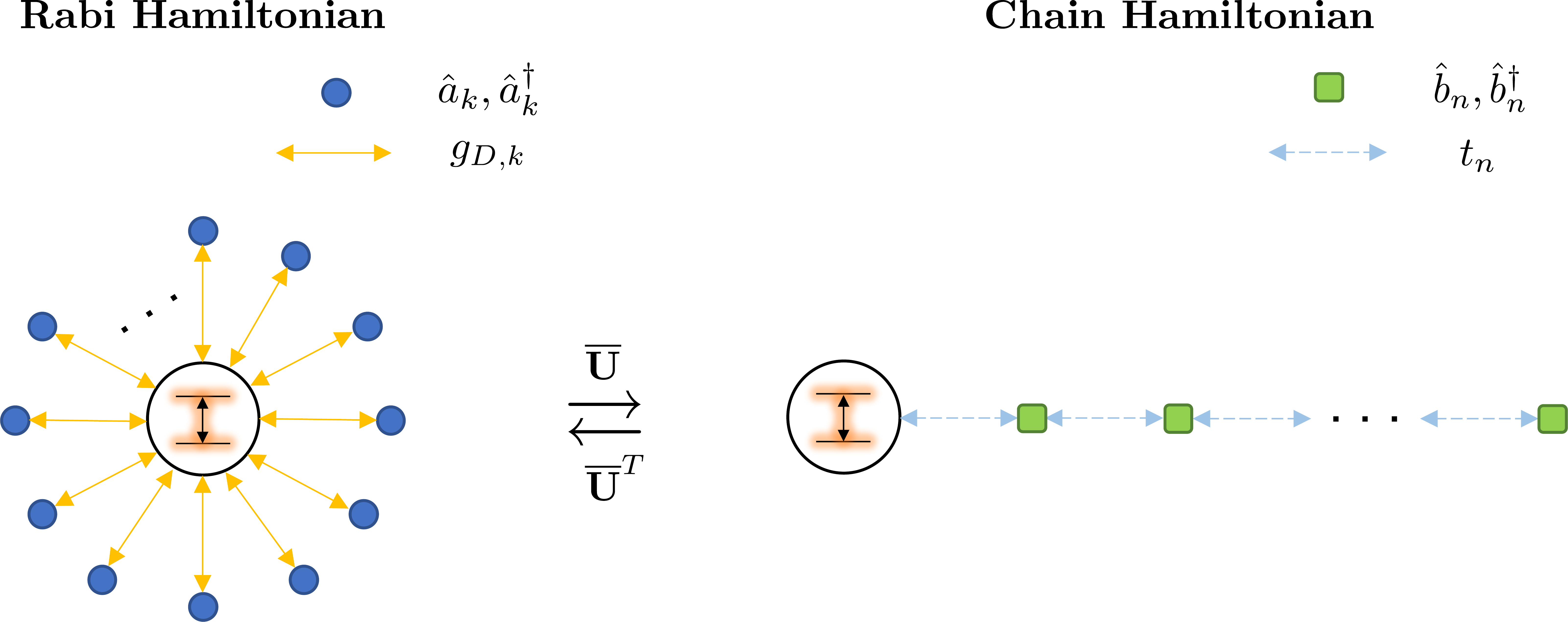}
\caption{The coupling structure of the TLA and the bosonic modes for the multimode Rabi Hamiltonian (left) and the chain Hamiltonian (right).}
\label{fig:MRH_to_chain}
\end{figure}

In order to simulate the multimode Rabi Hamiltonian (\ref{eq:Rabi_H_D}) using MPS, its coupling structure must be altered because, in its current form, the TLA is simultaneously coupled to all electromagnetic modes of the cavity. To make the coupling structure more compatible with MPS, the multimode Rabi Hamiltonian must be transformed to an equivalent Hamiltonian with a chain coupling structure as illustrated in Fig.\ \ref{fig:MRH_to_chain}. This transformation scheme has been derived for the spin-boson model \cite{Bulla_et_al_2005_NRG} and applied to the quantum impurity model \cite{Bulla_et_al_2008_NRG}. For the spin-boson model, analytical transformation schemes have been derived for linearly \cite{Prior_et_al_2010_Strong_system-environment_interactions,Chin_et_al_2010_exact_transformation_to_chain} and logarithmically \cite{Chin_et_al_2010_exact_transformation_to_chain} discretized bosonic modes, which represent bath oscillators. In this paper, we take the equations derived in \cite{Bulla_et_al_2005_NRG} and numerically solve them in a stable manner to implement the transformation to the chain Hamiltonian. This numerical transformation works for electromagnetic modes with arbitrarily distributed, discrete frequencies in contrast to the analytical schemes which only work for linearly and logarithmically distributed frequencies.

\subsection{General Transformation Scheme}\label{sec:general_transformation_scheme}
For a TLA in a cavity, the multimode Rabi Hamiltonian (\ref{eq:Rabi_H_D}) without the dipole self-energy term is
\begin{equation}\label{eq:Rabi_H_D_w/o_DSE}
\hat{\mathcal{H}}_D=\tfrac{\hbar\omega_a}{2}\hat\sigma_z+\hbar\sum_{k=1}^{M}\big[\omega_k\hat{a}_k^\dagger\hat{a}_k-i g_{D,k}\hat\sigma_x(\hat{a}_k-\hat{a}_k^\dagger)\big].
\end{equation}
Since the TLA is simultaneously coupled to all electromagnetic modes, the MPS implementation would be inefficient. For this reason, it is desirable to turn this Hamiltonian into one with 1D chain coupling structure with nearest-neighbor interactions.

The electromagnetic modes (or harmonic oscillators) represented by $\hat{a}_k$ and $\hat{a}_k^\dagger$ are associated with the vector potential spatial eigenfunctions ${\bf A}_k({\bf r})$ which form an orthonormal set of basis. It is thus possible to form a new set of orthonormal basis using a real, orthogonal matrix \cite{Bulla_et_al_2005_NRG}: $\hat{b}_n=\sum_{k=1}^{M}U_{n,k}\hat{a}_k$, where $[\overline{\bf U}]_{k,n}=U_{k,n}$, $\overline{\bf U}^T\cdot\overline{\bf U}=\overline{\bf U}\cdot\overline{\bf U}^T=\overline{\bf I}$, and $\overline{\bf I}$ is an $M\times M$ identity matrix. The inverse transform is naturally
\begin{equation}\label{eq:inverse_transform:chain_to_photonic}
\hat{a}_k=\sum_{n=1}^{M}U_{n,k}\hat{b}_n.
\end{equation}

There exists infinitely many orthogonal matrices for $\overline{\bf U}$, each of which implements a unique transformation on (\ref{eq:Rabi_H_D_w/o_DSE}) that results in an equivalent Hamiltonian with a different coupling structure. Observing that the coupling term in (\ref{eq:Rabi_H_D_w/o_DSE}) has the TLA coupled to a linear superposition of electromagnetic modes, we choose $U_{1,k}$ such that $\sum_k g_{D,k}\hat\sigma_x(\hat{a}_k-\hat{a}_k^\dagger)=\rho\hat\sigma_x(\hat{b}_1-\hat{b}_1^\dagger)$, where $\rho$ is a new coupling coefficient to be determined. This means the hybrid oscillator $\hat{b}_1$ is formed by lumping all the electromagnetic modes $\hat{a}_k$ together. This, in turn, induces the uncoupled harmonic oscillators to be coupled to each other in the transformed basis: $\sum_k \omega_k\hat{a}_k^\dagger\hat{a}_k=\sum_n[\xi_n\hat{b}_n^\dagger\hat{b}_n+t_n(\hat{b}_n^\dagger\hat{b}_{n+1}+\hat{b}_{n+1}^\dagger\hat{b}_n)]$ with a new set of bosonic mode frequencies $\xi_n$ and hopping parameters $t_n$. This proposed structure of the Hamiltonian gives rise to a recursive relation for finding rows of $\overline{\bf U}$ involving only the previous two rows of the matrix, which is simple to implement with low computational cost.

Combining what has been discussed above, (\ref{eq:Rabi_H_D_w/o_DSE}) is transformed into the chain Hamiltonian,
\begin{align}
\hat{\mathcal{H}}_\text{ch}&=\tfrac{\hbar\omega_a}{2}\hat\sigma_z-i\hbar\rho\hat\sigma_x(\hat{b}_1-\hat{b}_1^\dagger)\notag\\
&\quad+\hbar\sum_{n=1}^{M}\big[\xi_n\hat{b}_n^\dagger\hat{b}_n+t_n(\hat{b}_n^\dagger\hat{b}_{n+1}+\hat{b}_{n+1}^\dagger\hat{b}_n)\big],\label{eq:chain_Hamiltonian}
\end{align}
where $t_M=0$. The above is equivalent to (\ref{eq:Rabi_H_D_w/o_DSE}). For example, (\ref{eq:chain_Hamiltonian}) can be obtained by inserting (\ref{eq:inverse_transform:chain_to_photonic}) into (\ref{eq:Rabi_H_D_w/o_DSE}), and the other way is also possible by the inverse orthogonal transform. The remaining derivation of the transformation scheme is detailed in \cite{Bulla_et_al_2005_NRG}, and only the important results are summarized here. The rows of $\overline{\bf U}$ are recursively determined as
\begin{align}
U_{1,k}=\tfrac{g_{D,k}}{\rho},\quad U_{2,k}=\tfrac{\omega_k-\xi_1}{t_1}U_{1,k},\\
U_{n+1,k}=\tfrac{1}{t_n}[(\omega_k-\xi_n)U_{n,k}-t_{n-1}U_{n-1,k}]\label{eq:next_row_of_U}
\end{align}
with the coefficients,
\begin{subequations}
\begin{align}
\rho&=\sqrt{\textstyle\sum_{k=1}^{M}g_{D,k}^2},\\
t_1&=\tfrac{1}{\rho}\sqrt{\textstyle\sum_{k=1}^{M}(\omega_k-\xi_1)^2g_{D,k}^2},\\
\xi_n&=\textstyle\sum_{k=1}^{M}\omega_k U_{n,k}^2,\\
t_n&=\sqrt{\textstyle\sum_{k=1}^{M}[(\omega_k-\xi_n)U_{n,k}-t_{n-1}U_{n-1,k}]^2}.
\end{align}
\end{subequations}

\subsection{Stabilization of the Numerical Transformation}
\begin{figure}
\includegraphics[width=.9\linewidth]{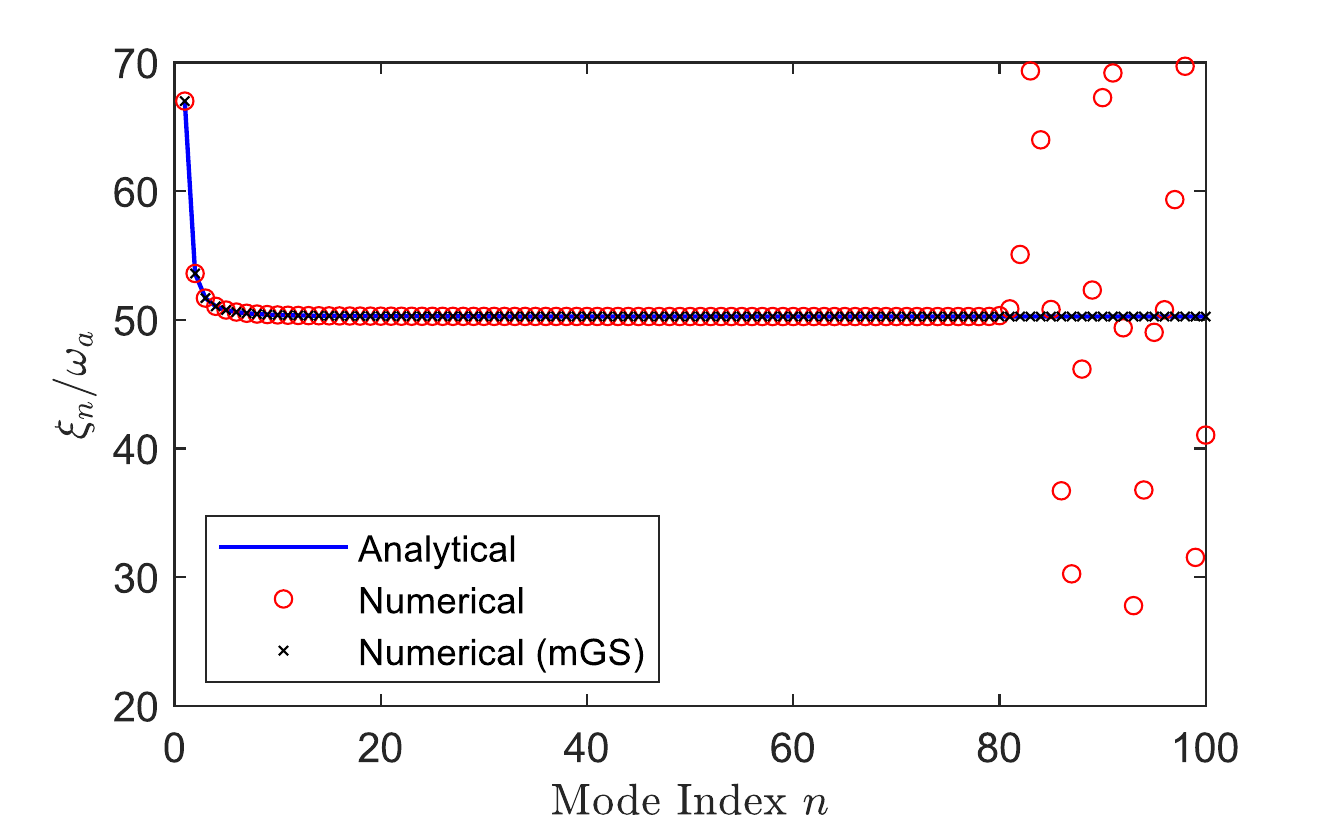}
\caption{Plots of normalized, transformed bosonic mode frequencies in the chain basis ($\xi_n/\omega_a$). While the straightforward numerical implementation suffers instability, the numerical scheme combined with modified Gram-Schmidt (mGS) shows good agreement with the analytical result.}
\label{fig:numerical_recursion_error}
\end{figure}

A straightforward numerical calculation of (\ref{eq:next_row_of_U}) leads to unstable solutions \cite{Bulla_et_al_2005_NRG,Bulla_et_al_2008_NRG}. Beyond what are suggested in \cite{Bulla_et_al_2008_NRG}, we have an effective remedy for this problem, which is to apply the modified Gram-Schmidt orthogonalization to every new row of $\overline{\bf U}$ calculated using (\ref{eq:next_row_of_U}). The accuracies of the numerical transformation schemes are shown in Fig.\ \ref{fig:numerical_recursion_error}. To test the schemes, an analytically solvable case of linear, discrete electromagnetic mode frequencies \cite{Chin_et_al_2010_exact_transformation_to_chain} is considered. The normalized, transformed mode frequencies in the chain basis ($\xi_n/\omega_a$) are plotted analytically, numerically, and numerically with modified Gram-Schmidt. The simple numerical scheme suffers instability, whereas the stabilized numerical scheme is completely accurate and stable as shown in Fig.\ \ref{fig:numerical_recursion_error}.

\section{Matrix Product States and Numerical Mode Decomposition}
\subsection{Time Evolution Using MPS}
\begin{figure}
\includegraphics[width=1\linewidth]{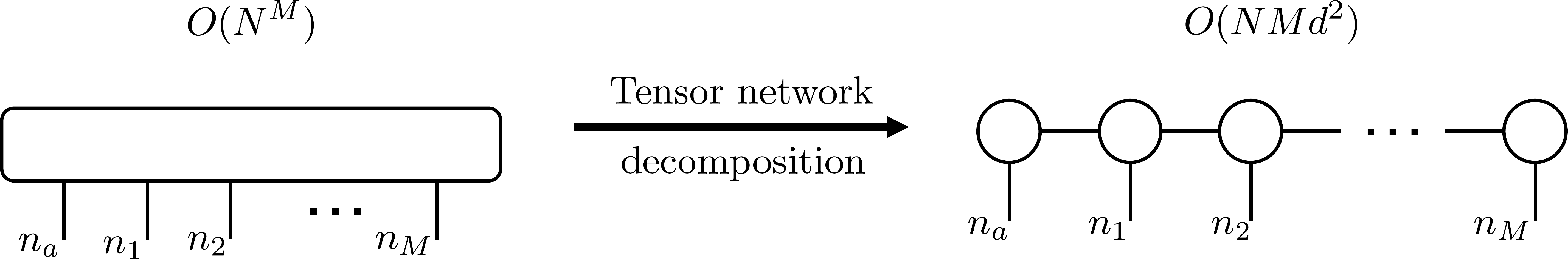}
\caption{Illustration of decomposing a high-order tensor (left) into an MPS (right).}
\label{fig:high-order_tensor_2_MPS}
\end{figure}

Once the chain Hamiltonian (\ref{eq:chain_Hamiltonian}) is obtained, the dynamics of the multimode Rabi Hamiltonian can be simulated efficiently by MPS, which is implemented using the tensor network contracting function provided in \cite{Pfeifer_et_al_2014_NCON}. A quantum state governed by the chain Hamiltonian is expressed in terms of a high-order tensor:
\begin{equation}\label{eq:high-order_quantum_state}
\ket{\psi_\text{ch}}=\sum_{\mathclap{n_a,n_1,\dots,n_M}}c_{n_a,n_1,\dots,n_M}\ket{n_a,n_1,\dots,n_M},
\end{equation}
where $n_a=0$ for $g$ and 1 for $e$, and $n_k$ with integer $k\in[1,M]$ is the Fock state for the $k$-th bosonic mode in the chain basis. When the number of photons is truncated for numerical calculations such that $n_k\in[0,N-1]$, (\ref{eq:high-order_quantum_state}) requires storing $O(N^M)$ complex numbers in the computer memory. By taking the tensor $c_{n_a,n_1,\dots,n_M}$ and decomposing it into $M+1$ tensors of order 3 or 2 (as illustrated in Fig.\ \ref{fig:high-order_tensor_2_MPS}), we obtain the MPS representation of the same state:
\begin{equation}\label{eq:MPS_rep_of_quantum_state}
\ket{\psi_\text{ch}}=\sum_{\mathclap{\substack{n_a,n_1,\dots,n_M\\a_1,\dots,a_{M}}}}A_{a_1}^{n_a}A_{a1,a2}^{n_1}\dots A_{a_{M-1},a_M}^{n_{M-1}}A_{a_M}^{n_M}\ket{n_a,n_1,\dots,n_M},
\end{equation}
where $A_{a_i,a_{i+1}}^{n_i}$ is an order 3 tensor that has three indices. This MPS representation now only requires storing $O(NMd^2)$ complex numbers, where $d$ is the maximum allowed bond dimension\footnote{The bond dimensions are the dimensions of indices $a_i$ and $a_{i+1}$ for an MPS tensor $A_{a_i,a_{i+1}}^{n_i}$. The bonds connect a tensor at a physical site with the neighboring tensors.} of the MPS.

The MPS representation (\ref{eq:MPS_rep_of_quantum_state}) can be made exactly equal to the original state (\ref{eq:high-order_quantum_state}), but it would not be efficient to do so. In practice, low-rank approximations  using singular value decomposition (SVD) are applied to reduce the bond dimensions between individual tensors of the MPS\@. To speed up this process, randomized SVD \cite{Halko_2011_rSVD} can be used. Consequently, (\ref{eq:MPS_rep_of_quantum_state}) becomes an approximate but much more efficient representation of (\ref{eq:high-order_quantum_state}).

Time evolving the quantum state boils down to solving the following time-stepping equation:
\begin{equation}\label{eq:time-stepping_eq}
\ket{\psi_\text{ch}(t+\Delta t)}=e^{-i\hat{\mathcal{H}}_\text{ch}\Delta t/\hbar}\ket{\psi_\text{ch}(t)}.
\end{equation}
To solve the above using MPS, the matrix product operator (MPO) representation of the time evolution operator must be formed first. This can be achieved using the time-evolving block decimation algorithm \cite{Paeckel_et_al2019_TE_methods_for_MPS}. This algorithm involves taking the chain Hamiltonian (\ref{eq:chain_Hamiltonian}) and dividing it into even and odd parts as $\hat{\mathcal{H}}_\text{ch}=\hat{\mathcal{H}}_\text{ch,even}+\hat{\mathcal{H}}_\text{ch,odd}$. Then, the time evolution operator is approximated:
\begin{equation}\label{eq:TEBD1}
e^{-i\hat{\mathcal{H}}_\text{ch}\Delta t/\hbar}\approx e^{-i\hat{\mathcal{H}}_\text{ch,even}\Delta t/\hbar}e^{-i\hat{\mathcal{H}}_\text{ch,odd}\Delta t/\hbar},
\end{equation}
where the approximate equality becomes exact in the limit $\Delta t\rightarrow0$. Because the terms within the even or odd part of the chain Hamiltonian commute, (\ref{eq:TEBD1}) can be readily turned into an MPO.

As the quantum state is evolved in time, the expectation values of various physical quantities can be computed to study the interaction dynamics of the system. Although all the information of the quantum dynamics is encoded into $\ket{\psi_\text{ch}(t)}$, which is represented by an MPS, it is necessary to construct the quantized field operators in order to study the behavior of the quantum fields emitted by the TLA.

\subsection{Field Quantization Using Mode Decomposition}\label{sec:field_quantization}
When the quantized vector potential operator in the Schr\"odinger picture is given by
\begin{equation}
\hat{\bf A}({\bf r})=\sum_{k=1}^{M}\sqrt{\tfrac{\hbar}{2\omega_k \epsilon_0 V_0}}\big[\hat{a}_k{\bf A}_k({\bf r})+\hat{a}_k^\dagger{\bf A}_k^*({\bf r})\big],
\end{equation}
the quantized electric field operator naturally follows as
\begin{equation}\label{eq:E-field_operator}
\hat{\bf E}({\bf r})=i\sum_{k=1}^{M}\sqrt{\tfrac{\hbar\omega_k}{2\epsilon_0 V_0}}\big[\hat{a}_k{\bf A}_k({\bf r})-\hat{a}_k^\dagger{\bf A}_k^*({\bf r})\big].
\end{equation}
The spatial eigenfunction ${\bf A}_k({\bf r})$ satisfies the vector wave equation,
\begin{equation}\label{eq:vector_wave_eqn}
\mu_0^{-1}\nabla\times\nabla\times{\bf A}_k({\bf r})-\omega_k^2\epsilon({\bf r}){\bf A}_k({\bf r})=0,
\end{equation}
and are properly normalized as
\begin{equation}
\frac{1}{V_0}\int_{V_0}dV\,{\bf A}_k^*({\bf r})\cdot\epsilon_r({\bf r}){\bf A}_{k'}({\bf r})=\delta_{kk'}
\end{equation}
with the relative permittivity $\epsilon_r({\bf r})=\epsilon({\bf r})/\epsilon_0$. The spatial eigenfunction is purely real for closed, perfect cavities. For certain cases where analytical solutions are available for (\ref{eq:vector_wave_eqn}), the analytical eigenfunctions are used to construct the quantized field operator. However, for a general, inhomogeneous case, numerical modes must be used. This technique of numerical mode decomposition (NMD) has been applied to simulate various quantum electromagnetic phenomena \cite{Qinfo_preserving_CEM,Na_Zhu_Chew_2021}.

\subsubsection*{Simulation Settings}
\begin{figure}
\includegraphics[width=1\linewidth]{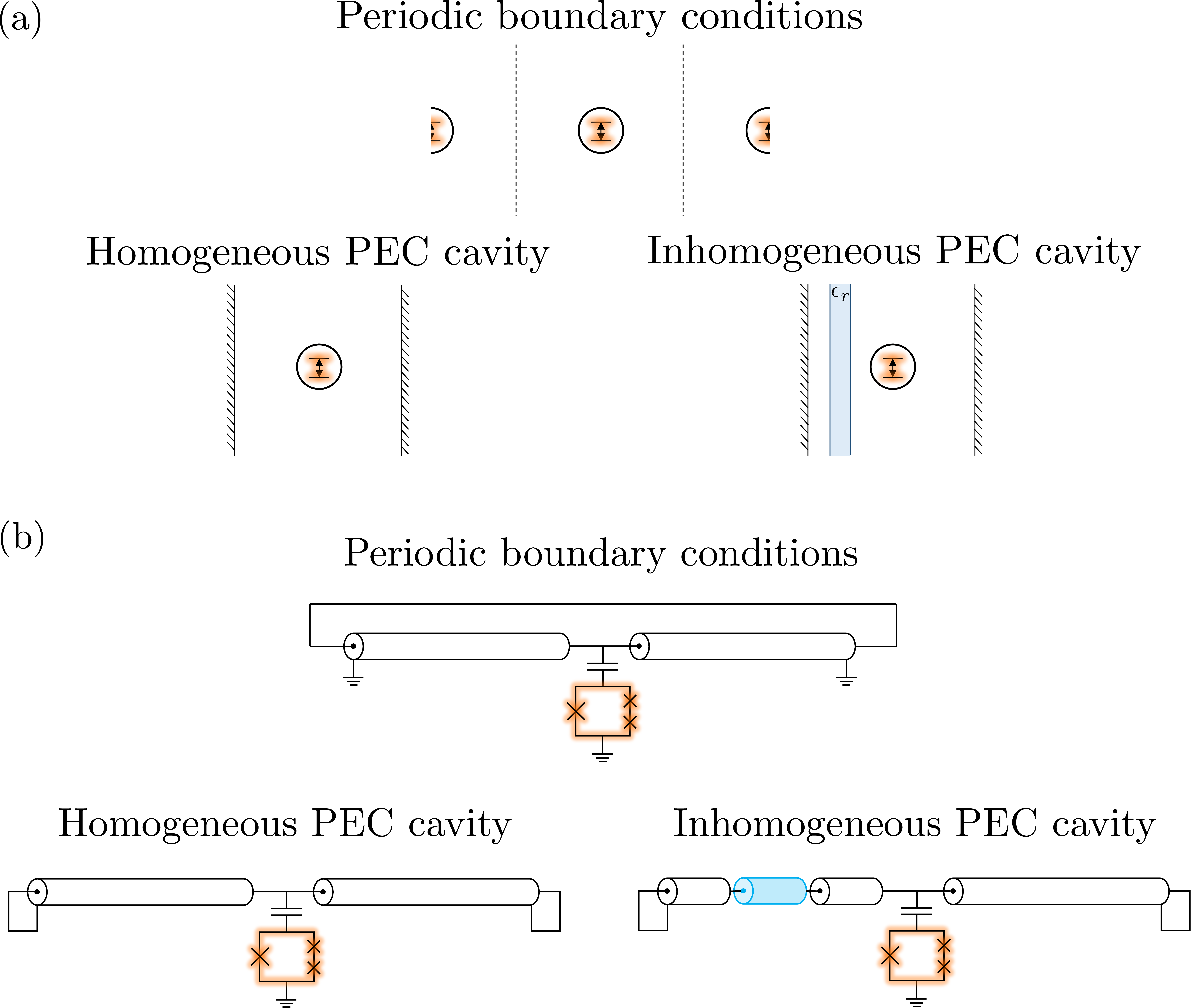}
\caption{(a) Illustration of the three different 1D simulation settings that are primarily considered in this paper. (b) Circuit analogues of the 1D simulation settings in (a). The components highlighted in orange are flux qubits implementing TLAs, capacitively coupled to transmission line resonators. The transmission line in solid blue color has a different characteristic impedance than the rest, causing discontinuous interfaces in the inhomogeneous case, but the junction effects in multi-section waveguides (such as mode conversion) are ignored in this work \cite{ChewECE604}.}
\label{fig:boundary_conditions_for_simulations}
\end{figure}

As a proof of concept, three kinds of simulation settings are considered in this paper: a TLA in a lattice with PBC, in a homogeneously filled PEC cavity,\footnote{This is an idealizing assumption that ignores any loss in the system. Therefore, the time-domain simulation results under this setting should be valid at time scales that are much shorter than the energy relaxation times for physical implementations of these QED systems.} and in an inhomogeneously filled PEC cavity, all of which are illustrated in Fig.\ \ref{fig:boundary_conditions_for_simulations}. Since these are 1D models that have variations only in the $x$-direction, they can be interpreted in terms of their circuit analogues also shown in Fig.\ \ref{fig:boundary_conditions_for_simulations}. The lattice/cavity is defined on $x\in[-L/2,L/2]$ where $L$ is its length. The TLA is placed at the center ($x=0$) for most simulations. The only nonzero component of the electric field is along the $z$-direction, and the dipole moment vector $\bf d$ of the TLA is perfectly aligned with the electric field.

\begin{figure}
\includegraphics[width=1\linewidth]{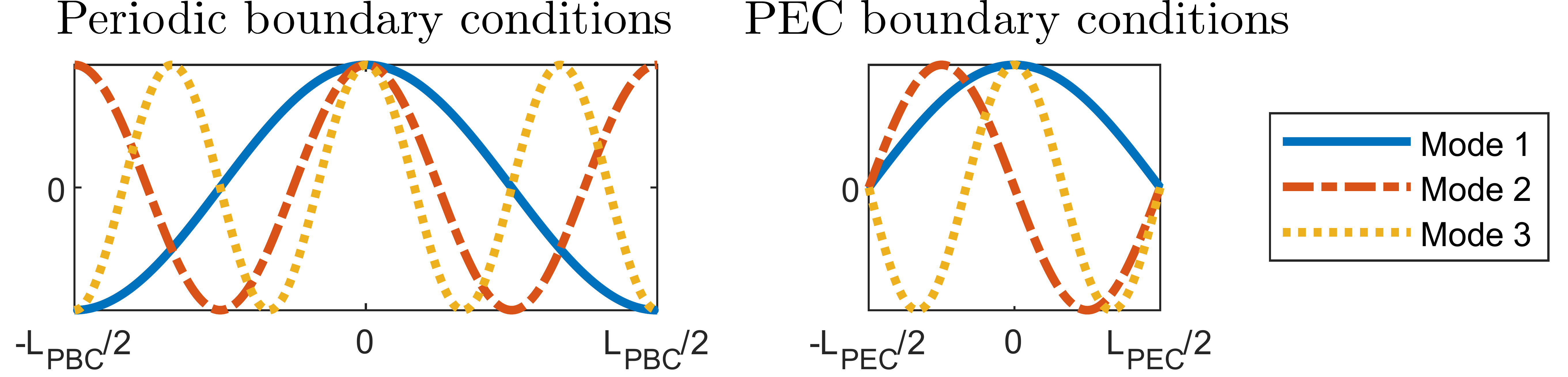}
\caption{1D electric field profiles for the first three modes with different boundary conditions. To have the lowest mode of the lattice/cavity resonant with the TLA, we have $L_\text{PBC}=2L_\text{PEC}=\lambda_a$.}
\label{fig:PBC_vs_PEC_mode_profiles}
\end{figure}

Moreover, the TLA is set to be resonant with the fundamental mode of the lattice/cavity. Because of this setting, the length of the lattice $L_\text{PBC}$ is different from the length of the cavity $L_\text{PEC}$ due to different boundary conditions. If we define the resonant wavelength in terms of the atomic frequency as $\lambda_a=2\pi c/\omega_a$, the lattice must be of length $L_\text{PBC}=\lambda_a$, while the cavity must be of length $L_\text{PEC}=\lambda_a/2$ (see Fig.\ \ref{fig:PBC_vs_PEC_mode_profiles}).

\subsubsection*{Analytical Modes}
In the PBC case, the solutions to (\ref{eq:vector_wave_eqn}) are ${\bf A}_k(x)={\bf e}_z e^{\pm ikx}$ where ${\bf e}_z$ is a unit vector in the $z$-direction. The wave number is $k=\omega_k/c$ where mode frequencies are $\omega_k=k\omega_a$ with integer $k\in[1,M]$.

The homogeneous PEC cavity case can also be handled analytically, and the eigenfunction in this case is ${\bf A}_k(x)={\bf e}_z\sqrt{2}\cos(kx)$. The mode frequencies are still $\omega_k=k\omega_a$ due to the PEC boundary conditions and the size of the cavity, $L_\text{PEC}=L_\text{PBC}/2$. However, when the TLA is placed at the center of the cavity, only the modes with odd $k$ couple to the TLA, so we denote the relevant modes in this case as $\omega_k=(2k-1)\omega_a$ with integer $k\in[1,M]$. The quantized electric field operator is constructed according to (\ref{eq:E-field_operator}) by inserting the proper eigenfunctions depending on the boundary conditions.

\subsubsection*{Numerical Modes}
For the general case of inhomogeneous PEC cavity, it is usually impossible to find analytical solutions to (\ref{eq:vector_wave_eqn}). Therefore, it must be solved by CEM techniques such as the finite element or finite difference method. By employing such a method, one discretizes the field and samples it at discrete points in 1D space:
\begin{equation}\label{eq:numerical_mode}
{\bf A}_k(j\Delta x)={\bf e}_z[\boldsymbol{\Phi}_k]_j
\end{equation}
where $\boldsymbol{\Phi}_k$ is an $N_x\times1$ vector for the $k$-th mode, the integer $j\in[1,N_x]$ indexes over the $N_x$ discrete points in space, and $\Delta x$ is the discretization length. With this, (\ref{eq:vector_wave_eqn}) is turned into a generalized eigenvalue problem of the form
\begin{equation}
\overline{\bf K}\cdot\boldsymbol{\Phi}_k=\omega_k^2\overline{\bf M}\cdot\boldsymbol{\Phi}_k,
\end{equation}
where $\overline{\bf K}$ is the stiffness matrix that implements a scaled, discrete curl-curl operation, $\overline{\bf M}$ is the mass matrix that describes the medium inhomogeneity $\epsilon(x)$, $\omega_k^2$ is the eigenvalue, and $\boldsymbol{\Phi}_k$ is the eigenvector. Once this is solved, the numerical modes (\ref{eq:numerical_mode}) are inserted into (\ref{eq:E-field_operator}) to construct the quantized electric field operator for the general inhomogeneous medium.

\section{Calculation of the Field Correlations Using Mode Decomposition}
With the quantum dynamics simulated using MPS and the field quantized by either analytical or numerical mode decomposition, we are ready to discuss the calculation of the first-order field correlation function.
\subsection{Formulation}
\subsubsection*{First-Order Field Correlation Function}
Given the quantized electric field operator (\ref{eq:E-field_operator}), its positive and negative frequency parts are
\begin{equation}
\hat{\bf E}^{(+)}({\bf r})=i\sum_{k=1}^{M}\sqrt{\tfrac{\hbar\omega_k}{2\epsilon_0 V_0}}\hat{a}_k{\bf A}_k({\bf r}),\quad\hat{\bf E}^{(-)}({\bf r})=[\hat{\bf E}^{(+)}({\bf r})]^\dagger.
\end{equation}
The quantity of interest is the first-order field correlation function,
\begin{widetext}
\begin{equation}
\braket{{\bf E}^{(-)}({\bf r})\cdot{\bf E}^{(+)}({\bf r})}=\tfrac{\hbar}{2\epsilon_0 V_0}\sum_{k,k'}\sqrt{\omega_k \omega_{k'}}{\bf A}_k^*({\bf r})\cdot{\bf A}_{k'}({\bf r})\braket{\psi_D(t)|\hat{a}_k^\dagger\hat{a}_{k'}|\psi_D(t)},\label{eq:first-order_correlation}
\end{equation}
where $\ket{\psi_D(t)}$ is obtained by time-evolving an initial state using the dipole gauge multimode Rabi Hamiltonian (\ref{eq:Rabi_H_D_w/o_DSE}) as $\ket{\psi_D(t)}=e^{-i\hat{\mathcal{H}}_D t/\hbar}\ket{\psi_{D,0}}$. The first-order field correlation function (\ref{eq:first-order_correlation}) represents the total counting rate of single photons at $\bf r$ (or average field intensity at that point) \cite{Walls_and_Milburn_Quantum_Optics,Gerry_and_Knight}.

However, the calculations with MPS are done in the chain basis by time evolution with the chain Hamiltonian (\ref{eq:chain_Hamiltonian}) as $\ket{\psi_\text{ch}(t)}=e^{-i\hat{\mathcal{H}}_\text{ch} t/\hbar}\ket{\psi_{\text{ch},0}}$. Since the transformation to the chain basis is realized with the orthogonal matrix $\overline{\bf U}$ as explained in Sec.\ \ref{sec:general_transformation_scheme}, the correlation function is reformulated in this basis:
\begin{equation}\label{eq:field_correlation_in_the_chain_basis}
\braket{{\bf E}^{(-)}({\bf r})\cdot{\bf E}^{(+)}({\bf r})}=\frac{\hbar}{2\epsilon_0 V_0}\sum_{k,k'}\sqrt{\omega_k \omega_{k'}}{\bf A}_k^*({\bf r})\cdot{\bf A}_{k'}({\bf r})\sum_{n,n'}U_{n,k}U_{n',k'}\braket{\psi_\text{ch}(t)|\hat{b}_n^\dagger\hat{b}_{n'}|\psi_\text{ch}(t)}.
\end{equation}

The numerical computation of (\ref{eq:field_correlation_in_the_chain_basis}) may seem burdensome since it involves four summations, each of which indexes through all $M$ modes. Nevertheless, the calculation of this quantity on a computer can be made more efficient by vectorizing the expression (for array programming). To do this, we define the correlation matrix in the chain basis as $[\overline{\bf B}]_{m,m'}=\braket{\psi_\text{ch}(t)|\hat{b}_m^\dagger\hat{b}_{m'}|\psi_\text{ch}(t)}$ and the columns of $\overline{\bf U}$ as $[{\bf u}_k]_n=U_{n,k}$. These lead to the vectorized form of (\ref{eq:field_correlation_in_the_chain_basis}), which can be computed efficiently:
\begin{align}
\braket{{\bf E}^{(-)}({\bf r})\cdot{\bf E}^{(+)}({\bf r})}=\tfrac{\hbar}{2\epsilon_0 V_0}\sum_{k,k'}\sqrt{\omega_k \omega_{k'}}{\bf A}_k^*({\bf r})\cdot{\bf A}_{k'}({\bf r})\left({\bf u}_k^T\cdot\overline{\bf B}\cdot{\bf u}_{k'}\right)&.
\end{align}
\end{widetext}

\subsubsection*{Average Number of Photons}
With the quantities defined above, it is simple to calculate the average number of photons in the $k$-th electromagnetic mode, $\braket{a_k^\dagger a_k}=\braket{\psi_D(t)|\hat{a}_k^\dagger\hat{a}_k|\psi_D(t)}$, which is converted to the chain basis and vectorized as
\begin{equation}
\braket{a_k^\dagger a_k}=\sum_{n,n'}U_{n,k}U_{n',k'}\braket{\psi_\text{ch}(t)|\hat{b}_n^\dagger\hat{b}_{n'}|\psi_\text{ch}(t)}={\bf u}_k^T\cdot\overline{\bf B}\cdot{\bf u}_k.
\end{equation}

\begin{figure*}
\includegraphics[width=1\linewidth]{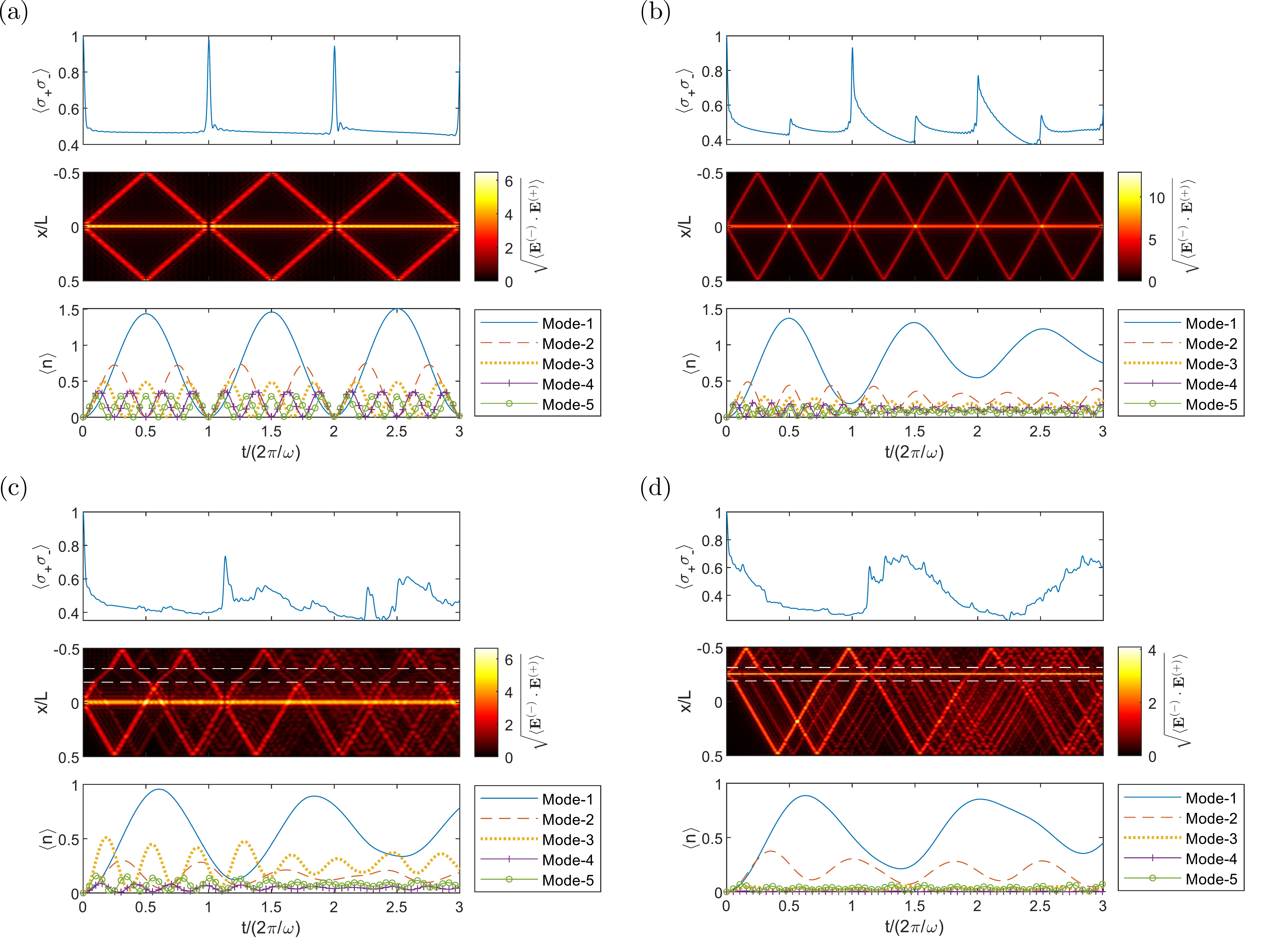}
\caption{Simulation of a TLA in various settings: (a) Periodic boundary conditions; (b) homogeneous PEC cavity; (c) TLA in a PEC cavity next to a dielectric slab; and (d) TLA in a dielectric slab in a PEC cavity. For each case, the three plots from top to bottom display the average population in the excited state of the TLA, first-order field correlation function in time and space, and average numbers of photons in five electromagnetic modes of lowest frequencies.}
\label{fig:MPS_sim_results}
\end{figure*}

\subsection{Simulation Results}
The simulation settings discussed in Sec.\ \ref{sec:field_quantization} are considered here, and three quantities are calculated: the average population in excited state of the TLA, first-order field correlation function, and average number of photons in each electromagnetic mode. These quantities are displayed in Fig.\ \ref{fig:MPS_sim_results} respectively. The latter two quantities are formulated in the previous subsection. The average population in the excited state is calculated as $\braket{\sigma_+\sigma_-}=\braket{\psi_\text{ch}(t)|\hat\sigma_+\hat\sigma_-|\psi_\text{ch}(t)}$, where the MPS representation of the state is defined in (\ref{eq:MPS_rep_of_quantum_state}). In all simulations, the quantum state is initialized to $\ket{\psi_{\text{ch},0}}=\ket{e,0,\dots,0}$, i.e., an excited TLA in vacuum. The coupling coefficient is tuned by the dipole moment vector $\bf d$ such that $g_{D,1}/\omega_1=0.6$, which represents ultrastrong coupling between the TLA and the fundamental mode of the cavity/lattice. This is a deliberate choice to be consistent with \cite{Munoz_superluminal_2018}.

Figure~\ref{fig:MPS_sim_results}(a) shows the simulation results for the PBC case, which is analyzed in \cite{Munoz_superluminal_2018}. The plots show a semi-periodic, resonant behavior of the TLA and fields. A lone TLA evolves in time as $e^{-i\omega_a t}$, so it has a period of $2\pi/\omega_a$. Light waves are emitted when the TLA decays, and the TLA is resonantly revived for a moment as it absorbs the waves that return at every period. When the emitted wave arrives at one end of the spatial simulation domain, it reappears at the opposite end due to PBC. This dynamics matches the results shown in \cite{Munoz_superluminal_2018}. The periodic revival behavior is not as perfect here because twenty modes are considered. The amplitude of the average number of photons in each mode is proportional to the normalized coupling coefficient squared, $(g_{D,k}/\omega_k)^{2}\propto1/\omega_k$, which has been noticed in \cite{Munoz_superluminal_2018} as well.

When the boundary conditions are changed to PEC, the behavior of the TLA changes completely as shown in Fig.~\ref{fig:MPS_sim_results}(b). The effect of the cavity size being halved ($L_\text{PEC}=L_\text{PBC}/2$ as explained in Sec.\ \ref{sec:field_quantization}) is clearly observed here. The light wave reflected from the walls of the cavity returns in half a period of the TLA [$t/(2\pi/\omega_a)=0.5$], so it is unable to absorb much of the energy, being out of phase with the fields. When the light returns again in a full period [$t/(2\pi/\omega_a)=1$], most of the energy is absorbed by the TLA. Due to the half-period excitations, the highly periodic behavior seen in the PBC case is lost in the PEC cavity case. However, the number of photons still is roughly proportional to $(g_{D,k}/\omega_k)^2$ as in the PBC case.

\begin{figure}
\includegraphics[width=.95\linewidth]{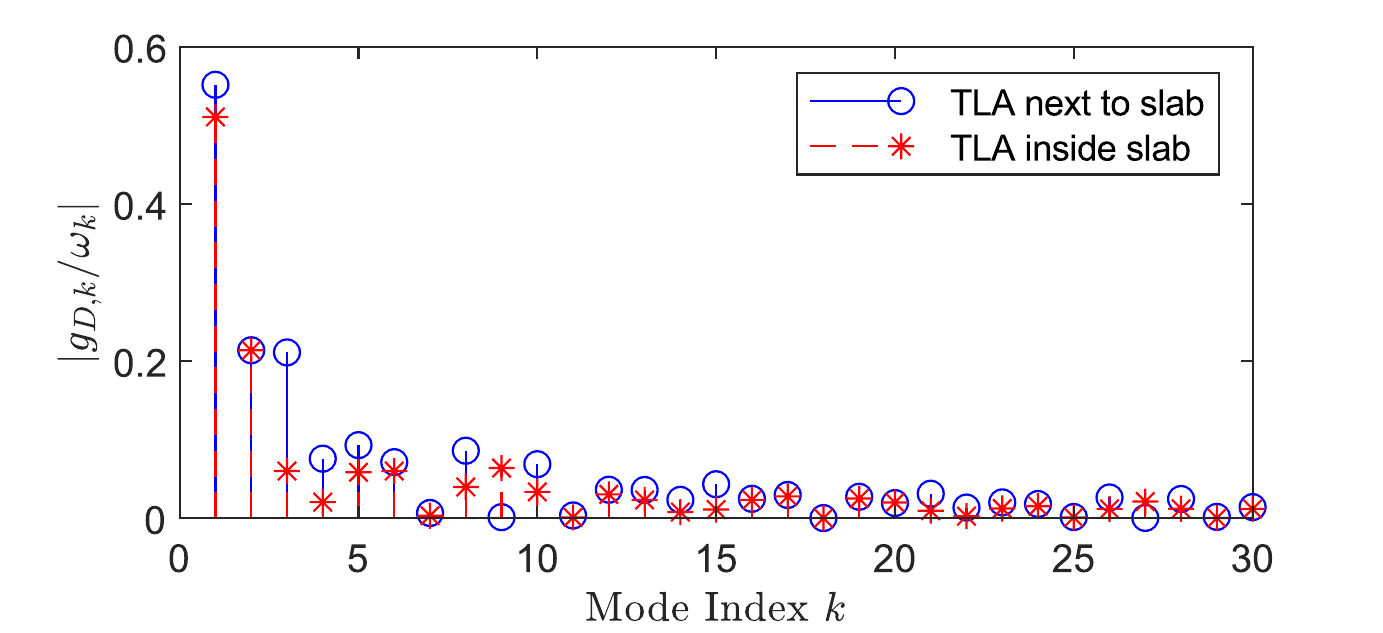}
\caption{Absolute values of normalized coupling coefficients for electromagnetic modes indexed by $k$. The TLA next to slab case corresponds to Fig.~\ref{fig:MPS_sim_results}(c), and the TLA inside slab case corresponds to Fig.~\ref{fig:MPS_sim_results}(d).}
\label{fig:inhom_coup_coef_plot}
\end{figure}

When a dielectric slab ($\epsilon_r=4$) of thickness $L_\text{PEC}/8$ is centered at $x=-L_\text{PEC}/4$, the dynamics changes drastically once again. In Figs.~\ref{fig:MPS_sim_results}(c) and \ref{fig:MPS_sim_results}(d), the dielectric slab is placed in between the white dashed lines shown in the field correlation plots. As a result of the accurate extraction of the numerical modes ${\bf A}_k(x)$ from NMD, all the reflections and transmissions at the vacuum-dielectric interfaces are properly realized. An interesting potential use case of our technique of MPS and NMD is demonstrated when the TLA is placed inside the dielectric slab [Fig.~\ref{fig:MPS_sim_results}(d)]. In this case, it is observed that photons of higher modes are strongly suppressed except for the first two modes. This can be understood when the normalized coupling coefficients shown in Fig.\ \ref{fig:inhom_coup_coef_plot} are observed. For most modes, the coefficients are smaller when the TLA is placed inside the slab. This directly results in less photons excited in these modes. Using this simulation approach, it is thus possible to design and engineer the electromagnetic environments for TLAs that allow dominant emission of photons for specific modes of the cavity. This could be useful when one seeks a design that produces monochromatic photons from a TLA source ultrastrongly coupled to its surrounding electromagnetic environment.

\section{Conclusion and Future Work}
We have derived the multimode Rabi Hamiltonians and identified the gauge invariant ones, confirming that the conclusions made with the single mode models \cite{Di_Stefano_et_al_2019_Resolution_of_gauge_ambiguities,Taylor_et_al_2020_Resolution_of_gauge_ambiguities} extend to the multimode case. We then chose to work with the dipole gauge multimode Rabi Hamiltonian to model cavity QED systems. To efficiently time step the quantum state, the Rabi Hamiltonian was transformed to the chain form, after which the MPS method was applied efficiently for time stepping the system. We have also presented the CEM technique of NMD, which allows modeling quantum electromagnetic fields not only for analytically solvable cases but also for general, inhomogeneous medium cases. The two methods, MPS and NMD, were combined to simulate the quantum dynamics of field-atom systems in various settings. The PBC setting was simulated to match the results with a previously published work \cite{Munoz_superluminal_2018}, and other more general cases were demonstrated to show the versatility of our approach to this problem.

Assuming the simple case of PEC cavity in this work is an idealizing assumption that ignores dissipation and loss which happens in the real world. This is a convenient assumption for our initial attempt at MPS simulations of cavity QED, but we have three ideas for more realistic models that can be considered for future work. First, bath oscillators can be added to the Hamiltonian \cite{Sha_Liu_Chew_DQEM_2018,Bruognolo_et_al_2014_Two-bath_SBM,Zueco_and_Garcia-Ripoll_2019_Ultrastrongly_dissipative_QRM}. Since the bath oscillators typically interact with the system weakly, it should be possible to incorporate them in MPS simulations of cavity QED\@. Second, an open cavity could be simulated by properly setting up the port boundary conditions \cite{Jin_EM_book,Roth_and_Braun_2023_Port_BC}. Third, quantization of the field using quasinormal modes would be another potential way to model open or dissipative cavity \cite{Sauvan_et_al_2013_Quasinormal_modes,Franke_et_al_2019_Quasinormal_modes}.

Our future work also includes extending this method to the multi-atom case and 3D problems. Extending the chain mapping technique (Sec.~\ref{sec:chain_mapping}) to multiple atoms is not trivial, but it would be absolutely essential for MPS simulations of quantum computing systems, where the effect of multiple qubits simultaneously interacting with multimode quantized fields is crucial. We also seek to numerically analyze practical 3D QED structures in the USC regime, such as a flux qubit coupled to a coplanar waveguide resonator \cite{Niemczyk_et_al_2010_Circuit_QED_in_the_USC_regime,Wang_et_al_2020_Bloch-Siegert_shift_in_US_coupled_circuit_QED}.

\begin{acknowledgments}
This work was supported by the National Science Foundation Grant No.~2202389 and teaching assistantship from the Department of Electrical and Computer Engineering at University of Illinois Urbana-Champaign.
\end{acknowledgments}

\appendix

\section{Minimal Coupling Hamiltonian}\label{app:min_coup_H}
The Hamiltonian for a free atom with a charge bound by the scalar potential is
\begin{equation}\label{eq:H_A}
\hat{H}_A=\frac{\hat{\bf p}^2}{2m}+q\Phi(\hat{\bf r}),
\end{equation}
where $m$ is the mass of the particle, $q$ is its electric charge, and the potential energy term is also written as $V(\hat{\bf r})=q\Phi(\hat{\bf r})$. When the atom is interacting with the electromagnetic fields in a cavity, the system must be invariant under a gauge transformation. The minimal coupling replacement,
\begin{equation}\label{eq:MCR}
\hat{\bf p}\rightarrow\hat{\bf p}-q\hat{\bf A}({\bf r}_0),
\end{equation}
makes the system gauge invariant. In the above, ${\bf r}_0$ is the position of the nucleus. Considering this replacement together with the energy in the electromagnetic fields of a cavity, the total Hamiltonian governing the system is
\begin{equation}\label{eq:H_C_not_expanded}
\hat{H}_C=\tfrac{[\hat{\bf p}-q\hat{\bf A}({\bf r}_0)]^2}{2m}+V(\hat{\bf r})+\tfrac{1}{2}\int dV\left(\tfrac{\hat{\boldsymbol{\Pi}}^2({\bf r})}{\epsilon({\bf r})}+\tfrac{[\nabla\times\hat{\bf A}({\bf r})]^2}{\mu_0}\right),
\end{equation}
which is called the minimal coupling Hamiltonian.

The integral term in (\ref{eq:H_C_not_expanded}) is the free field Hamiltonian that represents the total energy in the electromagnetic fields of a cavity:
\begin{equation}\label{eq:H_F}
\hat{H}_F=\frac{1}{2}\int dV\left(\frac{\hat{\boldsymbol{\Pi}}^2({\bf r})}{\epsilon({\bf r})}+\frac{[\nabla\times\hat{\bf A}({\bf r})]^2}{\mu_0}\right),
\end{equation}
where $\epsilon({\bf r})$ is the permittivity that describes the medium inhomogeneity, $\mu_0$ is the permeability of free space, and $\hat{\boldsymbol{\Pi}}({\bf r})=\epsilon({\bf r})\partial\hat{\bf A}({\bf r})/\partial t$ is the canonically conjugate variable of the vector potential operator $\hat{\bf A}$ \cite{QEM_a_new_look_I,QEM_a_new_look_II}. The above is in the Coulomb gauge ($\nabla\cdot\hat{\bf A}=0$), where the vector potential is purely transverse. The scalar potential $\Phi$ plays the role of the binding potential, $V(\hat{\bf r})=q\Phi(\hat{\bf r})$, for the free charge in the atom. For this reason, (\ref{eq:H_C_not_expanded}) is also called the \textit{Hamiltonian in the Coulomb gauge} \cite{Di_Stefano_et_al_2019_Resolution_of_gauge_ambiguities}. Considering a closed, perfect cavity, the number of modes is countably infinite, and (\ref{eq:H_F}) becomes $\hat{H}_F=\sum_{k=1}^{\infty}\hbar\omega_k\hat{a}_k^\dagger\hat{a}_k$, where $\omega_k$ is the frequency of the $k$-th cavity mode, $\hat{a}_k^\dagger$ ($\hat{a}_k$) is the photon creation (annihilation) operator for the $k$-th mode, and the zero-point energy term is ignored.

In \cite{Taylor_et_al_2022_Mode_truncation_cQED}, a proper way to truncate the electromagnetic modes has been defined by a projection operator of the form
\begin{equation}
\hat{P}^{(M)}=\hat{I}_A\otimes\left(\bigotimes_{k=1}^{M}\sum_{n_k=0}^{\infty}\ket{n_k}\bra{n_k}\bigotimes_{k'=M+1}^{\infty}\ket{0_{k'}}\bra{0_{k'}}\right),
\end{equation}
where $\hat{I}_A$ is the identity operator in the atomic Hilbert space, and $\ket{n_k}$ represents the Fock state of the $k$-th electromagnetic mode. This projection operator truncates the number of electromagnetic modes down to $M$ by allowing infinite numbers of photons in the modes with indices $k\in[1,M]$ and zero photons for all other modes. For example, the free field Hamiltonian would be truncated as
\begin{equation}\label{eq:H_F_truncated}
\hat{P}^{(M)}\hat{H}_F\hat{P}^{(M)}=\sum_{k=1}^{M}\hbar\omega_k\hat{a}_k^\dagger\hat{a}_k.
\end{equation}
Throughout the rest of this work, this proper truncation of electromagnetic modes is assumed.

\section{Coulomb and Dipole Gauge Hamiltonians}
Even though the Hamiltonian in the Coulomb gauge (\ref{eq:H_C_not_expanded}) is treated as the fundamental form by many, a gauge transformation is often applied to work in the dipole gauge where the resulting Hamiltonian is based on the electric dipole interaction. This process involves the Power-Zienau-Woolley (PZW) transformation \cite{Power_and_Zienau_1959_Coulomb_gauge_QED,Woolley_1971_Molecular_QED}. More recently, a simplified presentation of these Hamiltonians has been shown in \cite{Di_Stefano_et_al_2019_Resolution_of_gauge_ambiguities,Taylor_et_al_2020_Resolution_of_gauge_ambiguities}. This presentation and the equivalence of these Hamiltonians are shown in this section.

Many physicists prefer to work in the dipole gauge for various reasons. Some prefer it because they can avoid the potentials and directly work with the fields, and the Hamiltonian is manifestly gauge invariant \cite{Milonni}. It also turns out that the calculations involving electric dipole induced transitions in atoms are much more convenient to carry out in the dipole gauge \cite{Loudon_2000_Quantum_theory_of_light}. Another reason is that the canonical momentum ($\hat{\bf p}$) of the atom coincides with the kinematic momentum ($m\dot{\hat{\bf r}}$) in the dipole gauge \cite{Cohen-Tannoudji_Photons_and_Atoms}.

\subsection{Minimal Coupling Replacement as a Unitary Transformation}
The minimal coupling replacement (\ref{eq:MCR}) can be implemented as a unitary transformation \cite{Di_Stefano_et_al_2019_Resolution_of_gauge_ambiguities} with the operator
\begin{equation}\label{eq:U_MC}
\hat{U}_{MC}=e^{\frac{i}{\hbar}q\hat{\bf r}\cdot\hat{\bf A}({\bf r}_0)}.
\end{equation}
The Hamiltonian in (\ref{eq:H_C_not_expanded}) can then be written in terms of the minimal coupling replacement operation as
\begin{equation}\label{eq:MCH_with_MCR_operation}
\hat{H}_C=\hat{U}_{MC}\hat{H}_A\hat{U}_{MC}^\dagger+\hat{H}_F.
\end{equation}

To show that (\ref{eq:H_C_not_expanded}) can be obtained from (\ref{eq:MCH_with_MCR_operation}), the unitary transformation in the first term of the right-hand side of (\ref{eq:MCH_with_MCR_operation}) is evaluated. A lemma of the Baker-Campbell-Hausdorff (BCH) formula,
\begin{equation}\label{eq:BCH_formula}
\begin{aligned}
&e^{\hat{X}}\hat{Y}e^{-\hat{X}}\\
&=\hat{Y}+[\hat{X},\hat{Y}]+\tfrac{1}{2!}[\hat{X},[\hat{X},\hat{Y}]]+\tfrac{1}{3!}[\hat{X},[\hat{X},[\hat{X},\hat{Y}]]]+\dots,
\end{aligned}
\end{equation}
which is referred to as the BCH formula in the remainder of this paper, is extremely useful here. The above applies to any operators $\hat{X}$ and $\hat{Y}$ that may not commute, and the formula can be derived by applying Taylor expansions to the operator exponentials.

To evaluate the unitary transformation in the first term of (\ref{eq:MCH_with_MCR_operation}), we note that $\hat{H}_A$ (\ref{eq:H_A}) has two terms. The application of $\hat{U}_{MC}$ (\ref{eq:U_MC}) to the binding potential term of $\hat{H}_A$ is trivial because the term and unitary operator commute:
\begin{equation}
\hat{U}_{MC}V(\hat{\bf r})\hat{U}_{MC}^\dagger=e^{\frac{i}{\hbar}q\hat{\bf r}\cdot\hat{\bf A}({\bf r}_0)}V(\hat{\bf r})e^{-\frac{i}{\hbar}q\hat{\bf r}\cdot\hat{\bf A}({\bf r}_0)}=V(\hat{\bf r}).
\end{equation}
The momentum operator, on the other hand, changes under the transformation:
\begin{align}
\hat{U}_{MC}\hat{\bf p}\hat{U}_{MC}^\dagger&=e^{\frac{i}{\hbar}q\hat{\bf r}\cdot\hat{\bf A}({\bf r}_0)}\hat{\bf p}e^{-\frac{i}{\hbar}q\hat{\bf r}\cdot\hat{\bf A}({\bf r}_0)}\notag\\
&=\hat{\bf p}+\frac{i}{\hbar}q\sum_j [\hat{r}_j\hat{A}_j({\bf r}_0),\hat{\bf p}]\label{eq:MCR_on_p_intermediate_step}\\
&=\hat{\bf p}-q\hat{\bf A}({\bf r}_0),\notag
\end{align}
where the summation index goes through $j=x,y,$ and $z$, and the canonical commutation relation, $[\hat{r}_i,\hat{p}_j]=i\hbar\delta_{ij}\hat{I}$, has been used in the second and third equalities of the above. Also, the tensor product notation is omitted throughout the appendices.\footnote{The term $\hat{r}_j\hat{A}_j$ in (\ref{eq:MCR_on_p_intermediate_step}) is a shorthand notation for $\hat{r}_j\otimes\hat{A}_j$. The composite Hilbert space for the field-atom system is formed by a tensor product of the constituent Hilbert spaces: $\mathcal{H}=\mathcal{H}_A\otimes\mathcal{H}_F$. For this reason, the position operator $\hat{r}_j$ should technically be written as $\hat{r}_j\otimes\hat{I}_F$ where $\hat{I}_F$ is the identity operator in the field Hilbert space. Similarly, the vector potential operator should be $\hat{I}_A\otimes\hat{A}_j$ where $\hat{I}_A$ is the identity operator in the atomic Hilbert space. The position and vector potential operators clearly commute: $[\hat{r}_j\otimes\hat{I}_F,\hat{I}_A\otimes\hat{A}_j]=(\hat{r}_j\otimes\hat{I}_F)(\hat{I}_A\otimes\hat{A}_j)-(\hat{I}_A\otimes\hat{A}_j)(\hat{r}_j\otimes\hat{I}_F)=\hat{r}_j\otimes\hat{A}_j-\hat{r}_j\otimes\hat{A}_j=0$. Therefore, two operators that belong to different constituent Hilbert spaces always commute. For notational simplicity, the tensor products and identity operators are omitted throughout the appendices.} Since
\begin{equation}
\hat{U}_{MC}\hat{\bf p}^2\hat{U}_{MC}^\dagger=\hat{U}_{MC}\hat{\bf p}\hat{U}_{MC}^\dagger\hat{U}_{MC}\hat{\bf p}\hat{U}_{MC}^\dagger,
\end{equation}
the transformed atomic Hamiltonian is
\begin{equation}
\hat{U}_{MC}\hat{H}_A\hat{U}_{MC}^\dagger=\frac{[\hat{\bf p}-q\hat{\bf A}({\bf r}_0)]^2}{2m}+V(\hat{\bf r}).
\end{equation}
This reveals that (\ref{eq:MCH_with_MCR_operation}) is equal to (\ref{eq:H_C_not_expanded}). Finally, the Coulomb gauge Hamiltonian is written out:
\begin{equation}\label{eq:H_Capp}
\hat{H}_C=\frac{\hat{\bf p}^2}{2m}+V(\hat{\bf r})+\sum_{k=1}^{M} \hbar\omega_k \hat{a}_k^\dagger\hat{a}_k-\frac{q}{m}\hat{\bf p}\cdot\hat{\bf A}({\bf r}_0)+\frac{q^2}{2m}\hat{\bf A}^2({\bf r}_0).
\end{equation}

\subsection{Power-Zienau-Woolley Transformation}\label{app:PZW}
When the PZW transformation is applied to $\hat{H}_C$ given in (\ref{eq:MCH_with_MCR_operation}), the Hamiltonian in the dipole gauge is obtained. The unitary operator that implements the PZW transformation is given by
\begin{equation}\label{eq:U_PZW}
\hat{U}_{PZW}=\hat{U}_{MC}^\dagger=e^{-\frac{i}{\hbar}q\hat{\bf r}\cdot\hat{\bf A}({\bf r}_0)}.
\end{equation}
It is clearly seen in the above that $\hat{U}_{PZW}$ is just an inverse of $\hat{U}_{MC}$. This implies that the PZW transformation makes the replacement, $\hat{\bf p}\rightarrow\hat{\bf p}+q\hat{\bf A}({\bf r}_0)$, which is the opposite of the minimal coupling replacement. Later in this section, it is shown that the transformation also shifts the field operator. Applying this transformation to (\ref{eq:MCH_with_MCR_operation}) yields
\begin{align}
\hat{H}_D&=\hat{U}_{PZW}\hat{H}_C\hat{U}_{PZW}^\dagger\label{eq:H_D=U_H_C_U*}\\
&=\hat{U}_{PZW}\hat{U}_{MC}\hat{H}_A\hat{U}_{MC}^\dagger\hat{U}_{PZW}^\dagger+\hat{U}_{PZW}\hat{H}_F\hat{U}_{PZW}^\dagger\notag\\
&=\hat{H}_A+\hat{U}_{PZW}\hat{H}_F\hat{U}_{PZW}^\dagger,\label{eq:EDH_with_PZW_operation}
\end{align}
the Hamiltonian in the dipole gauge. It is observed in the above that the PZW transformation simply \textit{undoes} the minimal coupling replacement in the atomic part and shifts the field part of the Hamiltonian.

It is useful to write down the general form of the vector potential operator in the Schr\"odinger picture at this point:
\begin{equation}\label{eq:vec_pot_op}
\hat{\bf A}({\bf r})=\sum_{k=1}^{M}\sqrt{\frac{\hbar}{2\omega_k\epsilon_0 V_0}}\big[\hat{a}_k{\bf A}_k({\bf r})+\hat{a}_k^\dagger{\bf A}_k^*({\bf r})\big],
\end{equation}
where ${\bf A}_k({\bf r})$ is the vector potential eigenfunction. This operator is evaluated at ${\bf r}_0$ in (\ref{eq:U_PZW}) under the long-wavelength approximation. The vector potential eigenfunction, ${\bf A}_k({\bf r})$, is a real quantity for closed, perfect cavities, which are primarily dealt with in this work. Given (\ref{eq:vec_pot_op}), the quantized, transverse electric field operator naturally follows:
\begin{equation}\label{eq:transverse_E-field_operator}
\hat{\bf E}_\perp({\bf r})=i\sum_{k=1}^{M}\sqrt{\frac{\hbar\omega_k}{2\epsilon_0 V_0}}\big[\hat{a}_k{\bf A}_k({\bf r})-\hat{a}_k^\dagger{\bf A}_k^*({\bf r})\big].
\end{equation}

In order to specifically derive how the PZW transformation affects the free field Hamiltonian in (\ref{eq:EDH_with_PZW_operation}), the BCH formula (\ref{eq:BCH_formula}) is utilized here again. The transformation shifts the creation operator as
\begin{align}
&\hat{U}_{PZW}\hat{a}_k^\dagger\hat{U}_{PZW}^\dagger\notag\\
=&\,e^{-\frac{i}{\hbar}q\hat{\bf r}\cdot\hat{\bf A}({\bf r}_0)}\hat{a}_k^\dagger e^{\frac{i}{\hbar}q\hat{\bf r}\cdot\hat{\bf A}({\bf r}_0)}\notag\\
=&\,\hat{a}_k^\dagger-\frac{i}{\hbar}q\hat{\bf r}\cdot\sum_{k'=1}^{M}\sqrt{\frac{\hbar}{2\omega_k \epsilon_0 V_0}}{\bf A}_k({\bf r}_0)[\hat{a}_{k'},\hat{a}_k^\dagger]\notag\\
=&\,\hat{a}_k^\dagger-\frac{i}{\hbar}q\sqrt{\frac{\hbar}{2\omega_k \epsilon_0 V_0}}\hat{\bf r}\cdot{\bf A}_k({\bf r}_0)
\end{align}
where $[\hat{a}_{k'}^\dagger,\hat{a}_k^\dagger]=0$ has been used in the second equality of the above. Similarly, the annihilation operator is transformed as
\begin{equation}
\hat{U}_{PZW}\hat{a}_k\hat{U}_{PZW}^\dagger=\hat{a}_k+\frac{i}{\hbar}q\sqrt{\frac{\hbar}{2\omega_k \epsilon_0 V_0}}\hat{\bf r}\cdot{\bf A}_k^*({\bf r}_0).
\end{equation}
Considering ${\bf A}_k({\bf r}_0)$ to be real, we write
\begin{align}
&\,\hat{U}_{PZW}\hat{a}_k^\dagger\hat{a}_k\hat{U}_{PZW}^\dagger\notag\\
=&\,\hat{U}_{PZW}\hat{a}_k^\dagger\hat{U}_{PZW}^\dagger\hat{U}_{PZW}\hat{a}_k\hat{U}_{PZW}^\dagger\notag\\
=&\,\hat{a}_k^\dagger\hat{a}_k-i\frac{q\hat{\bf r}\cdot{\bf A}_k({\bf r}_0)(\hat{a}_k-\hat{a}_k^\dagger)}{\sqrt{2\hbar\omega_k \epsilon_0 V_0}}+\frac{q^2[\hat{\bf r}\cdot{\bf A}_k({\bf r}_0)]^2}{2\hbar\omega_k \epsilon_0 V_0},\label{eq:PZW_trans_applied_to_a*a}
\end{align}
where we identify the second term in (\ref{eq:PZW_trans_applied_to_a*a}) to be containing the $k$-th mode component of the transverse electric field operator (\ref{eq:transverse_E-field_operator}). Defining the dipole moment operator,
\begin{equation}\label{eq:dipole_moment_operator}
\hat{\bf d}=q\hat{\bf r},
\end{equation}
the free field Hamiltonian is transformed as
\begin{equation}\label{eq:U_PZW_H_F_U_PZW*}
\hat{U}_{PZW}\hat{H}_F\hat{U}_{PZW}^\dagger=\hat{H}_F-\hat{\bf d}\cdot\hat{\bf E}_\perp({\bf r}_0)+\sum_{k=1}^{M}\frac{[\hat{\bf d}\cdot{\bf A}_k({\bf r}_0)]^2}{2\epsilon_0V_0}.
\end{equation}
The last term in (\ref{eq:U_PZW_H_F_U_PZW*}) is called the dipole self-energy term. Therefore, the Hamiltonian in the dipole gauge is
\begin{equation}\label{eq:H_D_full}
\begin{aligned}
\hat{H}_D&=\frac{\hat{\bf p}^2}{2m}+V(\hat{\bf r})+\sum_{k=1}^{M} \hbar\omega_k\hat{a}_k^\dagger\hat{a}_k-\hat{\bf d}\cdot\hat{\bf E}_\perp({\bf r}_0)\\
&\quad+\sum_{k=1}^{M} \frac{[\hat{\bf d}\cdot{\bf A}_k({\bf r}_0)]^2}{2\epsilon_0 V_0},
\end{aligned}
\end{equation}
which is sometimes called the electric dipole Hamiltonian \cite{Milonni}.

It is obvious from (\ref{eq:H_D=U_H_C_U*}) that the Hamiltonians in the Coulomb and dipole gauges are equivalent because one is a unitary transformation of the other. This means they share the same energy eigenvalue spectrum, and therefore, they represent the same physical system. However, upon two-level truncation of the atomic part, the results from the two Hamiltonians do not agree anymore. This has been investigated for single electromagnetic mode models in \cite{Di_Stefano_et_al_2019_Resolution_of_gauge_ambiguities,Taylor_et_al_2020_Resolution_of_gauge_ambiguities,Stokes_and_Nazir_2020_Gauge_non-invariance} and resolved in \cite{Di_Stefano_et_al_2019_Resolution_of_gauge_ambiguities,Taylor_et_al_2020_Resolution_of_gauge_ambiguities}. We deal with the multimode case in the main text.

\section{Derivation of the Traditional Multimode Rabi Hamiltonians}
The Rabi Hamiltonian is obtained under the assumption that the frequency gap between the lowest two eigenstates of the atom is resonant with the fundamental electromagnetic mode of the cavity. In this case, the atomic part of the full Hamiltonian ($\hat{H}_C$ or $\hat{H}_D$) is truncated using the two-level projection operator:
\begin{equation}\label{eq:P-hat}
\hat{\mathcal{P}}=\ket{g}\bra{g}+\ket{e}\bra{e}
\end{equation}
where $\ket{g}$ and $\ket{e}$ are the ground and first excited states, respectively, of the free atomic Hamiltonian (\ref{eq:H_A}). When all of its eigenstates and energy eigenvalues are found, this Hamiltonian can be diagonalized as
\begin{equation}\label{eq:H_A_diag}
\hat{H}_A=\sum_n E_n\ket{E_n}\bra{E_n}.
\end{equation}
In terms of these eigenstates, $\ket{g}=\ket{E_0}$ and $\ket{e}=\ket{E_1}$. A useful relation between the projection and identity operators is
\begin{equation}\label{eq:I_A=P+Q}
\hat{I}_A=\hat{\mathcal{P}}+\hat{\mathcal{Q}}
\end{equation}
where $\hat{\mathcal{Q}}=\sum_{n\ge2} \ket{E_n}\bra{E_n}$.

Using the two-level projection operator (\ref{eq:P-hat}), the traditional way to obtain the Rabi Hamiltonian is through a direct truncation:
\begin{equation}\label{eq:direct_truncation}
\hat{\mathcal{H}}_i'=\hat{\mathcal{P}}\hat{H}_i\hat{\mathcal{P}}
\end{equation}
where $i=C$ or $D$ for the Coulomb or dipole gauge. Such way of truncation applied in both gauges is demonstrated in \cite{Leonardi1986}. The prime on the left-hand side of the above indicates a direct truncation as opposed to a more proper way of truncation which is discussed in Appendix~\ref{app:gauge_invar_Rabi_H}. Calligraphic symbols with hats are used to represent the operators in the truncated two-level subspace of the atomic Hilbert space. Equations (\ref{eq:Rabi_H_C'}) and (\ref{eq:Rabi_H_D'}) are derived in this section.

\subsection{Direct Truncation of the Dipole Gauge Hamiltonian}
The truncation (\ref{eq:direct_truncation}) is applied to (\ref{eq:H_D_full}) as
\begin{equation}\label{eq:Rabi_H_D'app}
\hat{\mathcal{H}}_D'=\hat{\mathcal{P}}\hat{H}_A\hat{\mathcal{P}}+\hat{H}_F-\hat{\mathcal{P}}\hat{\bf d}\hat{\mathcal{P}}\cdot\hat{\bf E}_\perp({\bf r}_0)+\sum_{k=1}^{M}\frac{\hat{\mathcal{P}}[\hat{\bf d}\cdot{\bf A}_0({\bf r}_0)]^2\hat{\mathcal{P}}}{2\epsilon_0 V_0},
\end{equation}
where the two-level truncation on the field Hamiltonian (second term in the above) is omitted for notational simplicity.\footnote{As discussed in the previous footnote, we do not denote identity operators and tensor products for notational simplicity. The field Hamiltonian should technically be written as $\hat{I}_A\otimes\hat{H}_F$ before the two-level truncation and as $(\hat{\mathcal{P}}\otimes\hat{I}_F)(\hat{I}_A\otimes\hat{H}_F)(\hat{\mathcal{P}}\otimes\hat{I}_F)=\hat{\mathcal{P}}\otimes\hat{H}_F=\hat{\mathcal{I}}\otimes\hat{H}_F$ after, where $\hat{\mathcal{I}}=\hat{\mathcal{P}}$ is an identity operator in the truncated two-level subspace. We omit the $\hat{\mathcal{I}}\,\otimes$ part since it consists of an identity operator and a tensor product.}

The two-level truncation of the first term of (\ref{eq:Rabi_H_D'app}) represents the truncation of the free atomic Hamiltonian. Using (\ref{eq:P-hat}) and (\ref{eq:H_A_diag}),
\begin{align}\label{eq:two-level_atomic_H_incl_I}
\hat{\mathcal{P}}\hat{H}_A\hat{\mathcal{P}}&=E_0\ket{g}\bra{g}+E_1\ket{e}\bra{e}\notag\\
&=\tfrac{E_1-E_0}{2}(\ket{e}\bra{e}-\ket{g}\bra{g})+\tfrac{E_1+E_0}{2}(\ket{e}\bra{e}+\ket{g}\bra{g})\notag\\
&=\tfrac{\hbar\omega_a}{2}\hat{\sigma}_z+\tfrac{E_1+E_0}{2}\hat{\mathcal{I}},
\end{align}
where $\hbar\omega_a=E_1-E_0$, and
\begin{equation}\label{eq:I_in_two-lvl_subspace}
\hat{\mathcal{I}}=\hat{\mathcal{P}}
\end{equation}
is the identity operator in the two-level subspace. In (\ref{eq:two-level_atomic_H_incl_I}), the second term is usually dropped from the Rabi Hamiltonian because it is proportional to the identity operator.

The third term in (\ref{eq:Rabi_H_D'app}) is the interaction term. For symmetric potentials centered at the origin in space, $\hat{\mathcal{P}}\hat{\bf d}\hat{\mathcal{P}}$ only has nonzero elements in the off-diagonal entries. The dipole moment operator in the two-level subspace is written as
\begin{equation}
\hat{\mathcal{P}}\hat{\bf d}\hat{\mathcal{P}}={\bf d}\hat{\sigma}_-+{\bf d}^*\hat{\sigma}_+
\end{equation}
where ${\bf d}=\braket{g|\hat{\bf d}|e}$ is the complex dipole moment vector. By properly choosing the phases of the eigenstates $\ket{g}$ and $\ket{e}$, the dipole moment $\bf d$ can be made real \cite{DABMiller}. What is meant here is that the phase of an eigenstate is arbitrary: for example, it is possible to make the replacement $\ket{g}\rightarrow e^{-i\phi_g}\ket{g}$ without altering the orthogonality of the eigenstates. Therefore, by properly selecting the phases of the eigenstates, the dipole moment vector can be computed as (assuming a $z$-directed dipole for simplicity)
\begin{align}
{\bf d}&={\bf e}_z q\braket{g|e^{i(\phi_g-\phi_e)}\hat{z}|e}\notag\\
&={\bf e}_z qe^{i(\phi_g-\phi_e)}\int dz\,z\psi_g^*(z)\psi_e(z).
\end{align}
It is always possible to pick the phase difference $\phi_g-\phi_e$ such that the above quantity is real. Therefore, the third term of (\ref{eq:Rabi_H_D'app}) is written out as
\begin{equation}\label{eq:electric_dipole_interaction_term}
-\hat{\mathcal{P}}\hat{\bf d}\hat{\mathcal{P}}\cdot\hat{\bf E}_\perp({\bf r}_0)=-i\hbar\sum_{k=1}^{M}\underbrace{{\bf d}\cdot{\bf A}_k({\bf r}_0)\sqrt{\tfrac{\omega_k}{2\hbar\epsilon_0 V_0}}}_{=g_{D,k}}\hat{\sigma}_x(\hat{a}_k-\hat{a}_k^\dagger),
\end{equation}
where $\hat\sigma_x=\hat\sigma_-+\hat\sigma_+$ is a Pauli operator, and $g_{D,k}$ is the mode-dependent coupling coefficient in the dipole gauge.

The last term of (\ref{eq:Rabi_H_D'app}) is a diagonal operator in the atomic Hilbert space, and it does not include any field operator. For this reason, this term (the dipole self-energy term) only modifies the atomic energy and does not alter the interaction \cite{Loudon_2000_Quantum_theory_of_light}.

Gathering the results, the multimode Rabi Hamiltonian in the dipole gauge obtained by a direct truncation is
\begin{equation}\label{eq:Rabi_H_D'_evaluated}
\begin{aligned}
\hat{\mathcal{H}}_D'=\frac{\hbar\omega_a}{2}\hat{\sigma}_z+\sum_{k=1}^{M}\Big[&\hbar\omega_k\hat{a}_k^\dagger\hat{a}_k-i\hbar g_{D,k}\hat{\sigma}_x(\hat{a}_k-\hat{a}_k^\dagger)\\
&+\frac{1}{2\epsilon_0 V}\hat{\mathcal{P}}[\hat{\bf d}\cdot{\bf A}_k({\bf r}_0)]^2\hat{\mathcal{P}}\Big].
\end{aligned}
\end{equation}

\subsection{Direct Truncation of the Coulomb Gauge Hamiltonian}
When the direct truncation (\ref{eq:direct_truncation}) is applied to the Coulomb gauge Hamiltonian (\ref{eq:H_Capp}), gauge invariance is ruined (reasons are discussed in Appendix~\ref{app:trunc_method_and_gauge_invar}), and the following results:
\begin{align}
\hat{\mathcal{H}}_C'&=\hat{\mathcal{P}}\hat{U}_{MC}\hat{H}_A\hat{U}_{MC}^\dagger\hat{\mathcal{P}}+\hat{H}_F\label{eq:Rabi_H_C'_high-level}\\
&=\hat{\mathcal{P}}\hat{H}_A\hat{\mathcal{P}}-\frac{q}{m}\hat{\mathcal{P}}\hat{\bf p}\hat{\mathcal{P}}\cdot\hat{\bf A}({\bf r}_0)+\frac{q^2}{2m}\hat{\bf A}^2({\bf r}_0)+\hat{H}_F.\label{eq:Rabi_H_C'_not_evaluted_yet}
\end{align}
The first term of the above has already been evaluated in (\ref{eq:two-level_atomic_H_incl_I}). In order to apply the two-level truncation to the interaction term, the third term in (\ref{eq:Rabi_H_C'_not_evaluted_yet}), as explained in \cite{Leonardi1986}, the relation between the position and momentum operators must be established in the two-level subspace. To this end, the free atomic Hamiltonian (\ref{eq:H_A}) is considered here again. The Heisenberg equation of motion for the position operator implies the following relation:
\begin{equation}\label{eq:p=[r,H]}
\hat{\bf p}=\frac{m}{i\hbar}[\hat{\bf r},\hat{H}_A].
\end{equation}
Expanding the commutator and taking the matrix element of the operators on both sides yields
\begin{equation}
\braket{g|\hat{\bf p}|e}=-\frac{im}{q\hbar}(\underbrace{E_1-E_0}_{=\hbar\omega_a})\underbrace{\braket{g|q\hat{\bf r}|e}}_{={\bf d}}
\end{equation}
where the dipole moment vector has been identified. This equation relates the matrix element of the momentum operator to the electric dipole moment. Using this result, the interaction term is written in the two-level subspace as
\begin{equation}\label{eq:minimal_coupling_interaction_term}
-\frac{q}{m}\hat{\mathcal{P}}\hat{\bf p}\hat{\mathcal{P}}\cdot\hat{\bf A}({\bf r}_0)=\hbar\sum_{k=1}^{M}\underbrace{\frac{\omega_a{\bf d}\cdot{\bf A}_k({\bf r}_0)}{\sqrt{2\hbar\omega_k \epsilon_0 V_0}}}_{=g_{C,k}}\hat{\sigma}_y(\hat{a}_k+\hat{a}_k^\dagger),
\end{equation}
where $\hat\sigma_y=i(\hat\sigma_--\hat\sigma_+)$ is a Pauli operator, and the mode-dependent coupling coefficient in the Coulomb gauge ($g_{C,k}$) is identified above.

It is interesting to compare the coupling coefficients in the two gauges from (\ref{eq:electric_dipole_interaction_term}) and (\ref{eq:minimal_coupling_interaction_term}). They are related to each other as
\begin{equation}\label{eq:g_C=g_D*w_a/w_f}
g_{C,k}=g_{D,k}\frac{\omega_a}{\omega_k}.
\end{equation}
This is why the Rabi Hamiltonians in the two gauges appear equivalent on resonance for single mode models \cite{Scully_and_Zubairy_1997_QO}.

Therefore, the Rabi Hamiltonian in the Coulomb gauge when truncated directly is written out as
\begin{equation}\label{eq:Rabi_H_C'app}
\begin{aligned}
\hat{\mathcal{H}}_C'&=\frac{\hbar\omega_a}{2}\hat\sigma_z+\hbar\sum_{k=1}^{M}\left[\omega_k\hat{a}_k^\dagger\hat{a}_k+g_{C,k}\hat\sigma_y(\hat{a}_k+\hat{a}_k^\dagger)\right]\\
&\quad+\frac{\hbar}{\omega_a}\Big[\sum_{k=1}^{M}g_{C,k}(\hat{a}_k+\hat{a}_k^\dagger)\Big]^2.
\end{aligned}
\end{equation}
The coefficient of the last term in the above, called the diamagnetic term, is obtained by applying the Thomas-Reiche-Kuhn (TRK) sum rule as detailed in Appendix~\ref{app:TRK_sum_rule}\@. Although the Coulomb and dipole gauge Hamiltonians ($\hat{H}_C$ and $\hat{H}_D$) are equivalent, the Rabi Hamiltonians with direct truncation ($\hat{\mathcal{H}}_C'$ and $\hat{\mathcal{H}}_D'$) are not. Gauge invariance has been lost in the process of direct two-level truncation.

\section{Derivation of the Properly Truncated Multimode Rabi Hamiltonians}\label{app:gauge_invar_Rabi_H}
The two Rabi Hamiltonians derived in the previous section are not equivalent because after the truncation, the Hamiltonians are not unitarily related anymore. There is, however, a more proper way to truncate the Hamiltonian such that the unitarity between the Hamiltonians in both gauges is maintained \cite{Di_Stefano_et_al_2019_Resolution_of_gauge_ambiguities,Taylor_et_al_2020_Resolution_of_gauge_ambiguities}. When the full Hamiltonian (before the truncation) is thought of as a function of four conjugate operators, $\hat{H}_i(\hat{\bf r},\hat{\bf p},\hat{\bf A},\hat{\boldsymbol{\Pi}})$, the following truncation method keeps unitarity between the two gauges:
\begin{equation}\label{eq:prop_trunc}
\hat{\mathcal{H}}_i=\hat{H}_i(\hat{\mathcal{P}}\hat{\bf r}\hat{\mathcal{P}},\hat{\mathcal{P}}\hat{\bf p}\hat{\mathcal{P}},\hat{\bf A},\hat{\boldsymbol{\Pi}})
\end{equation}
where the subscript $i=C$ or $D$ for the Coulomb or dipole gauge. Using this proper truncation method, Eqs.~(\ref{eq:Rabi_H_C}) and (\ref{eq:Rabi_H_D}) from the main text are derived in this section.

\begin{widetext}
\subsection{Truncation of the Coulomb Gauge Hamiltonian}
The initial form of the properly truncated Coulomb gauge Hamiltonian is written from (\ref{eq:prop_trunc}) and (\ref{eq:MCH_with_MCR_operation}) as
\begin{align}
\hat{\mathcal{H}}_C&=\hat{H}_C(\hat{\mathcal{P}}\hat{\bf r}\hat{\mathcal{P}},\hat{\mathcal{P}}\hat{\bf p}\hat{\mathcal{P}},\hat{\bf A},\hat{\boldsymbol{\Pi}})\notag\\
&=e^{iq\hat{\mathcal{P}}\hat{\bf r}\hat{\mathcal{P}}\cdot\hat{\bf A}({\bf r}_0)/\hbar}\hat{H}_A(\hat{\mathcal{P}}\hat{\bf r}\hat{\mathcal{P}},\hat{\mathcal{P}}\hat{\bf p}\hat{\mathcal{P}})e^{-iq\hat{\mathcal{P}}\hat{\bf r}\hat{\mathcal{P}}\cdot\hat{\bf A}({\bf r}_0)/\hbar}+\hat{H}_F\notag\\
&=\hat{\mathcal{U}}_{MC}\hat{\mathcal{H}}_A\hat{\mathcal{U}}_{MC}^\dagger+\hat{H}_F,\label{eq:Rabi_H_C_high-level_form}
\end{align}
where
\begin{equation}
\hat{\mathcal{U}}_{MC}=e^{\frac{i}{\hbar}q\hat{\mathcal{P}}\hat{\bf r}\hat{\mathcal{P}}\cdot\hat{\bf A}({\bf r}_0)}=e^{i\sum_{k=1}^{M}\frac{g_{D,k}}{\omega_k}\hat\sigma_x(\hat{a}_k+\hat{a}_k^\dagger)}
\end{equation}
is the minimal coupling replacement operator in the two-level subspace, and
\begin{equation}
\hat{\mathcal{H}}_A=\frac{\hbar\omega_a}{2}\hat\sigma_z
\end{equation}
is the free two-level atomic Hamiltonian derived in (\ref{eq:two-level_atomic_H_incl_I}).

To evaluate the first term of (\ref{eq:Rabi_H_C_high-level_form}), the BCH formula (\ref{eq:BCH_formula}) comes in handy here. Applying it yields
\begin{align}
\hat{\mathcal{U}}_{MC}\hat\sigma_z\hat{\mathcal{U}}_{MC}^\dagger&=\hat\sigma_z+i\textstyle\sum_{k=1}^{M}\tfrac{g_{D,k}}{\omega_k}(\hat{a}_k+\hat{a}_k^\dagger)[\hat\sigma_x,\hat\sigma_z]+\frac{\big[i\textstyle\sum_{k=1}^{M}\tfrac{g_{D,k}}{\omega_k}(\hat{a}_k+\hat{a}_k^\dagger)\big]^2}{2!}[\hat\sigma_x,[\hat\sigma_x,\hat\sigma_z]]+\dots\notag\\ 
&=\hat\sigma_z+2\textstyle\sum_{k=1}^{M}\tfrac{g_{D,k}}{\omega_k}(\hat{a}_k+\hat{a}_k^\dagger)\hat\sigma_y-\frac{\big[2\textstyle\sum_{k=1}^{M}\tfrac{g_{D,k}}{\omega_k}(\hat{a}_k+\hat{a}_k^\dagger)\big]^2}{2!}\hat\sigma_z-\dots\notag\\
&=\hat\sigma_z\cos\big[\textstyle\sum_{k=1}^{M}\tfrac{2g_{D,k}}{\omega_k}(\hat{a}_k+\hat{a}_k^\dagger)\big]+\hat\sigma_y\sin\big[\textstyle\sum_{k=1}^{M}\tfrac{2g_{D,k}}{\omega_k}(\hat{a}_k+\hat{a}_k^\dagger)\big]
\end{align}
where the commutation relations for the Pauli operators ($[\hat\sigma_j,\hat\sigma_k]=2i\varepsilon_{jkl}\hat\sigma_l$ with the Levi-Civita symbol $\varepsilon_{jkl}$) and the Taylor series expansions for sine and cosine have been used. Therefore, the properly truncated Rabi Hamiltonian in the Coulomb gauge is
\begin{equation}\label{eq:Rabi_H_Capp}
\hat{\mathcal{H}}_C=\sum_{k=1}^{M}\hbar\omega_k\hat{a}_k^\dagger\hat{a}_k+\frac{\hbar\omega_a}{2}\left\{\hat\sigma_z\cos\big[\textstyle\sum_{k=1}^{M}\tfrac{2g_{D,k}}{\omega_k}(\hat{a}_k+\hat{a}_k^\dagger)\big]+\hat\sigma_y\sin\big[\textstyle\sum_{k=1}^{M}\tfrac{2g_{D,k}}{\omega_k}(\hat{a}_k+\hat{a}_k^\dagger)\big]\right\}
\end{equation}
which is the multimode version of Equation (10) in \cite{Di_Stefano_et_al_2019_Resolution_of_gauge_ambiguities}.

\subsection{Truncation of the Dipole Gauge Hamiltonian}
In a similar manner to how (\ref{eq:Rabi_H_C_high-level_form}) is written down, we can do the same in the dipole gauge using (\ref{eq:prop_trunc}) and (\ref{eq:EDH_with_PZW_operation}):
\begin{align}
\hat{\mathcal{H}}_D&=\hat{H}_D(\hat{\mathcal{P}}\hat{\bf r}\hat{\mathcal{P}},\hat{\mathcal{P}}\hat{\bf p}\hat{\mathcal{P}},\hat{\bf A},\hat{\boldsymbol{\Pi}})\notag\\
&=\hat{H}_A(\hat{\mathcal{P}}\hat{\bf r}\hat{\mathcal{P}},\hat{\mathcal{P}}\hat{\bf p}\hat{\mathcal{P}})+e^{-iq\hat{\mathcal{P}}\hat{\bf r}\hat{\mathcal{P}}\cdot\hat{\bf A}({\bf r}_0)/\hbar}\hat{H}_F e^{iq\hat{\mathcal{P}}\hat{\bf r}\hat{\mathcal{P}}\cdot\hat{\bf A}({\bf r}_0)/\hbar}\notag\\
&=\hat{\mathcal{H}}_A+\hat{\mathcal{U}}_{PZW}\hat{H}_F\hat{\mathcal{U}}_{PZW}^\dagger\label{eq:Rabi_H_D_high-level}
\end{align}
where we identify the PZW transformation operator in the two-level subspace,
\begin{equation}\label{eq:U_PZW_in_two-lvl_subspace}
\hat{\mathcal{U}}_{PZW}=\hat{\mathcal{U}}_{MC}^\dagger=e^{-i\sum_{k=1}^{M}\frac{g_{D,k}}{\omega_k}\hat\sigma_x(\hat{a}_k+\hat{a}_k^\dagger)}.
\end{equation}
This operator can be used to transform the Coulomb gauge Rabi Hamiltonian (\ref{eq:Rabi_H_C}) as
\begin{equation}
\hat{\mathcal{H}}_D=\hat{\mathcal{U}}_{PZW}\hat{\mathcal{H}}_C\hat{\mathcal{U}}_{PZW}^\dagger.\label{eq:unitary_Rabi_Hamiltonians}
\end{equation}

The transformation of the field Hamiltonian, which is the second term in (\ref{eq:Rabi_H_D_high-level}), is evaluated using the BCH formula (\ref{eq:BCH_formula}) again in a very similar manner as it is done in Appendix~\ref{app:PZW}. The PZW transformation in the two-level subspace (\ref{eq:U_PZW_in_two-lvl_subspace}) is applied to the creation and annihilation operators:
\begin{align}
\hat{\mathcal{U}}_{PZW}\hat{a}_k^\dagger\hat{\mathcal{U}}_{PZW}^\dagger&=\hat{a}_k^\dagger-i\sum_{k'=1}^{M}\tfrac{g_{D,k'}}{\omega_{k'}}\hat\sigma_x[\hat{a}_{k'}+\hat{a}_{k'}^\dagger,\hat{a}_k^\dagger]=\hat{a}_k^\dagger-i\tfrac{g_{D,k}}{\omega_k}\hat\sigma_x,\\
\hat{\mathcal{U}}_{PZW}\hat{a}_k\hat{\mathcal{U}}_{PZW}^\dagger&=\hat{a}_k-i\sum_{k'=1}^{M}\tfrac{g_{D,k'}}{\omega_{k'}}\hat\sigma_x[\hat{a}_{k'}+\hat{a}_{k'}^\dagger,\hat{a}_k]=\hat{a}_k+i\tfrac{g_{D,k}}{\omega_k}\hat\sigma_x.
\end{align}
Using the fact that
\begin{equation}
\hat{\mathcal{U}}_{PZW}\hat{a}_k^\dagger\hat{a}_k\hat{\mathcal{U}}_{PZW}^\dagger=\hat{\mathcal{U}}_{PZW}\hat{a}_k^\dagger\hat{\mathcal{U}}_{PZW}^\dagger\hat{\mathcal{U}}_{PZW}\hat{a}_k\hat{\mathcal{U}}_{PZW}^\dagger,
\end{equation}
the free field Hamiltonian is transformed as
\begin{align}
\hat{\mathcal{U}}_{PZW}\hat{H}_F\hat{\mathcal{U}}_{PZW}^\dagger&=\sum_{k=1}^{M}\hbar\omega_k(\hat{a}_k^\dagger-i\tfrac{g_{D,k}}{\omega_k}\hat\sigma_x)(\hat{a}_k+i\tfrac{g_{D,k}}{\omega_k}\hat\sigma_x)\notag\\
&=\sum_{k=1}^{M}\hbar\omega_k\big[\hat{a}_k^\dagger\hat{a}_k+\tfrac{g_{D,k}^2}{\omega_k^2}\hat\sigma_x^2-i\tfrac{g_{D,k}}{\omega_k}\hat\sigma_x(\hat{a}_k-\hat{a}_k^\dagger)\big]\notag\\
&=\sum_{k=1}^{M}\left[\hbar\omega_k\hat{a}_k^\dagger\hat{a}_k+\hbar g_{D,k}\hat\sigma_y(\hat{a}_k-\hat{a}_k^\dagger)+\hbar\tfrac{g_{D,k}^2}{\omega_k}\hat{\mathcal{I}}\right]
\end{align}
where $\hat\sigma_x^2=\hat{\mathcal{I}}$, and $\hat{\mathcal{I}}$ has been defined in (\ref{eq:I_in_two-lvl_subspace}).

Therefore, the properly truncated Rabi Hamiltonian in the dipole gauge is
\begin{equation}\label{eq:Rabi_H_Dapp}
\hat{\mathcal{H}}_D=\frac{\hbar\omega_a}{2}\hat\sigma_z+\hbar\sum_{k=1}^{M}\left[\omega_k\hat{a}_k^\dagger\hat{a}_k-ig_{D,k}\hat\sigma_x(\hat{a}_k-\hat{a}_k^\dagger)+\tfrac{g_{D,k}^2}{\omega_k}\hat{\mathcal{I}}\right].
\end{equation}
The coefficient of the dipole self-energy term (last term in the above) in the two-level subspace is derived in Appendix~\ref{app:TRK_sum_rule}.

\section{Truncation Method and Unitarity}\label{app:trunc_method_and_gauge_invar}
Other than the reasons mentioned in \cite{Di_Stefano_et_al_2019_Resolution_of_gauge_ambiguities,Taylor_et_al_2020_Resolution_of_gauge_ambiguities} about why gauge invariance is ruined when the direct truncation method is applied, it is also possible to mathematically see that this is the case because unitarity between the two gauges is lost. The directly truncated Rabi Hamiltonians are
\begin{align}
\hat{\mathcal{H}}_C'&=\hat{\mathcal{P}}\hat{H}_C\hat{\mathcal{P}}=\hat{\mathcal{P}}e^{iq\hat{\bf r}\cdot\hat{\bf A}({\bf r}_0)/\hbar}\hat{H}_A(\hat{\bf r},\hat{\bf p})e^{-iq\hat{\bf r}\cdot\hat{\bf A}({\bf r}_0)/\hbar}\hat{\mathcal{P}}+\hat{H}_F,\\
\hat{\mathcal{H}}_D'&=\hat{\mathcal{P}}\hat{H}_D\hat{\mathcal{P}}=\hat{\mathcal{P}}\hat{H}_A(\hat{\bf r},\hat{\bf p})\hat{\mathcal{P}}+\hat{\mathcal{P}}e^{-iq\hat{\bf r}\cdot\hat{\bf A}({\bf r}_0)/\hbar}\hat{H}_F e^{iq\hat{\bf r}\cdot\hat{\bf A}({\bf r}_0)/\hbar}\hat{\mathcal{P}}.
\end{align}
The above two are not unitarily related partly because $\hat{\mathcal{P}}e^{iq\hat{\bf r}\cdot\hat{\bf A}({\bf r}_0)/\hbar}$ is not unitary:
\begin{align}
\hat{\mathcal{P}}e^{iq\hat{\bf r}\cdot\hat{\bf A}({\bf r}_0)/\hbar}(\hat{\mathcal{P}}e^{iq\hat{\bf r}\cdot\hat{\bf A}({\bf r}_0)/\hbar})^\dagger&=\hat{\mathcal{P}},\\
(\hat{\mathcal{P}}e^{iq\hat{\bf r}\cdot\hat{\bf A}({\bf r}_0)/\hbar})^\dagger \hat{\mathcal{P}}e^{iq\hat{\bf r}\cdot\hat{\bf A}({\bf r}_0)/\hbar}&=e^{-iq\hat{\bf r}\cdot\hat{\bf A}({\bf r}_0)/\hbar}\hat{\mathcal{P}}e^{iq\hat{\bf r}\cdot\hat{\bf A}({\bf r}_0)/\hbar}.
\end{align}
In other words, there exists no unitary operator $\hat{\mathcal{U}}$ such that $\hat{\mathcal{H}}_D'=\hat{\mathcal{U}}\hat{\mathcal{H}}_C'\hat{\mathcal{U}}^\dagger$.

In contrast, the properly truncated Rabi Hamiltonians are clearly unitarily related:
\begin{align}
\hat{\mathcal{H}}_C&=\hat{H}_C(\hat{\mathcal{P}}\hat{\bf r}\hat{\mathcal{P}},\hat{\mathcal{P}}\hat{\bf p}\hat{\mathcal{P}},\hat{\bf A},\hat{\boldsymbol{\Pi}})=e^{iq\hat{\mathcal{P}}\hat{\bf r}\hat{\mathcal{P}}\cdot\hat{\bf A}({\bf r}_0)/\hbar}\hat{H}_A(\hat{\mathcal{P}}\hat{\bf r}\hat{\mathcal{P}},\hat{\mathcal{P}}\hat{\bf p}\hat{\mathcal{P}})e^{-iq\hat{\mathcal{P}}\hat{\bf r}\hat{\mathcal{P}}\cdot\hat{\bf A}({\bf r}_0)/\hbar}+\hat{H}_F,\\
\hat{\mathcal{H}}_D&=\hat{H}_D(\hat{\mathcal{P}}\hat{\bf r}\hat{\mathcal{P}},\hat{\mathcal{P}}\hat{\bf p}\hat{\mathcal{P}},\hat{\bf A},\hat{\boldsymbol{\Pi}})=\hat{H}_A(\hat{\mathcal{P}}\hat{\bf r}\hat{\mathcal{P}},\hat{\mathcal{P}}\hat{\bf p}\hat{\mathcal{P}})+e^{-iq\hat{\mathcal{P}}\hat{\bf r}\hat{\mathcal{P}}\cdot\hat{\bf A}({\bf r}_0)/\hbar}\hat{H}_F e^{iq\hat{\mathcal{P}}\hat{\bf r}\hat{\mathcal{P}}\cdot\hat{\bf A}({\bf r}_0)/\hbar}.
\end{align}
They are unitary transforms of each other by the PZW operator in the two-level subspace:
\begin{equation}
\hat{\mathcal{U}}_{PZW}=e^{-iq\hat{\mathcal{P}}\hat{\bf r}\hat{\mathcal{P}}\cdot\hat{\bf A}({\bf r}_0)/\hbar}.
\end{equation}
Thus, it is possible to write $\hat{\mathcal{H}}_D=\hat{\mathcal{U}}_{PZW}\hat{\mathcal{H}}_C\hat{\mathcal{U}}_{PZW}^\dagger$, and this means the two Rabi Hamiltonians share the same spectra and represent the same physical system.
\end{widetext}

\section{Application of the Thomas-Reiche-Kuhn Sum Rule}\label{app:TRK_sum_rule}
The TRK sum rule \cite{Reiche_and_Thomas_1925_Sum_rule,Kuhn_1925_Sum_rule} is a useful identity for atomic electrons derived from the canonical commutation relation. It has been used to find the coefficient of the diamagnetic term of the Rabi Hamiltonian in the Coulomb gauge \cite{Tufarelli_et_al2015_A2_term}. We derive this and the coefficient of the dipole self-energy term in the dipole gauge for the multimode case here.

\subsection{TRK Sum Rule}
Given the free atomic Hamiltonian (\ref{eq:H_A}), the TRK sum rule is written from \cite{Wang_1999_Generalization_of_TRK_sum_rule} as
\begin{equation}\label{eq:TRK_r}
\sum_n (E_n-E_0)|\braket{E_0|\hat{\bf r}|E_n}|^2=\frac{3\hbar^2}{2m}.
\end{equation}
This is a useful relation because multiplying both sides by the electric charge squared results in the weighted sum of the transition dipole moment equaling a constant. The generalized version of the TRK sum rule for any Hermitian observable $\hat{O}$ (as derived in \cite{Wang_1999_Generalization_of_TRK_sum_rule}) is
\begin{equation}\label{eq:TRK_general}
\sum_n (E_n-E_0)|\braket{E_0|\hat{O}|E_n}|^2=\frac{1}{2}\braket{E_0|[\hat{O},[\hat{O},\hat{H}_A]]|E_0}.
\end{equation}
This relation can be derived by applying the resolution of the identity and the fact that $E_n$ are the energy eigenvalues of the free atomic Hamiltonian.

\subsection{Coefficient of the Diamagnetic Term}
The coefficient of the diamagnetic term is derived using the TRK sum rule in \cite{Tufarelli_et_al2015_A2_term}. Their result can be easily extended to the multimode case as
\begin{equation}
\tfrac{q^2}{2m}\hat{\bf A}^2\ge\sum_n\tfrac{|\braket{E_n|\frac{q}{m}\hat{\bf p}\cdot\hat{\bf A}|E_0}|^2}{E_n-E_0}=\tfrac{\hbar}{\omega_a}\Big[\sum_{k=1}^{M}g_{C,k}(\hat{a}_k+\hat{a}_k^\dagger)\Big]^2
\end{equation}
where the inequality is saturated when $\hat{\bf p}$ is perfectly aligned with $\hat{\bf A}$.

\subsection{Coefficient of the Dipole Self-Energy Term}
The same technique can be applied for the dipole self-energy term of $\hat{\mathcal{H}}_D$ (\ref{eq:Rabi_H_D}) that is proportional to $\hat{\bf d}^2$ in the dipole gauge Hamiltonian. This is called the TRK sum rule for interacting photons \cite{Savasta_et_al_2021_TRK_sum_rule}, and the following equation can be obtained:
\begin{equation}
\frac{[\hat{\bf d}\cdot{\bf A}_k({\bf r}_0)]^2}{2\epsilon_0 V}=\sum_{n_k}\frac{|\braket{n_k|\hat{\bf d}\cdot\hat{\bf E}_{\perp,k}({\bf r}_0)|0_k}|^2}{\hbar\omega_k n_k},
\end{equation}
where $\ket{n_k}$ is the $k$-th mode Fock state, and $\hat{\bf E}_{\perp,k}$ is the transverse electric field operator for the $k$-th mode. Under two-level truncation, the above reduces to
\begin{equation}\label{eq:d-squared_coef}
\begin{aligned}
\frac{[\hat{\mathcal{P}}\hat{\bf d}\hat{\mathcal{P}}\cdot{\bf A}_k({\bf r}_0)]^2}{2\epsilon_0 V}&=\sum_{n_k}\frac{|\braket{n_k|\hat{\mathcal{P}}\hat{\bf d}\hat{\mathcal{P}}\cdot\hat{\bf E}_{\perp,k}({\bf r}_0)|0_k}|^2}{\hbar\omega_k n_k}\\
&=\tfrac{\hbar g_{D,k}^2}{\omega_k}\hat\sigma_x^2.
\end{aligned}
\end{equation}
Considering $\hat\sigma_x^2=\hat{\mathcal{I}}=\hat{\mathcal{P}}$ is the identity operator in the truncated two-level subspace, the above confirms the coefficient of the dipole self-energy term in Eq.~(\ref{eq:Rabi_H_D}) of the main text.

\end{document}